\documentclass[fleqn,usenatbib]{mnras}
\usepackage{newtxtext,newtxmath}
\usepackage[T1]{fontenc}

\DeclareRobustCommand{\VAN}[3]{#2}
\let\VANthebibliography\thebibliography
\def\thebibliography{\DeclareRobustCommand{\VAN}[3]{##3}\VANthebibliography}

\usepackage{breqn}
\usepackage{graphicx}	

\usepackage{multirow}
\usepackage{balance}
\usepackage{geometry}
\usepackage[usenames,dvipsnames]{xcolor}
\hypersetup{
    colorlinks = true,
    citecolor = {MidnightBlue},
    linkcolor = {BrickRed},
    urlcolor = {BrickRed}
}

\title[Statistical recovery of the BAO scale]{Statistical recovery of the BAO scale from multipoles of the beam-convolved 21cm correlation function}

\author[F.~Kennedy \& P.~Bull]{
Fraser Kennedy$^{1}$\thanks{E-mail: f.kennedy@qmul.ac.uk}
and
Philip Bull$^{1,2}$
\\
$^{1}$Astronomy Unit, Queen Mary University of London, Mile End Road, London E1 4NS, United Kingdom \\
$^{2}$Department of Physics and Astronomy, University of Western Cape, Cape Town 7535, South Africa
}

\date{Accepted XXX. Received YYY; in original form ZZZ}

\pubyear{2020}

\begin{document}
\label{firstpage}
\pagerange{\pageref{firstpage}--\pageref{lastpage}}
\maketitle

\begin{abstract}
Despite being designed as an interferometer, the MeerKAT radio array (an SKAO pathfinder) can also be used in autocorrelation (`single-dish') mode, where each dish scans the sky independently. Operating in this mode allows extremely high survey speeds to be achieved, albeit at significantly lower angular resolution. We investigate the recovery of the baryon acoustic oscillation (BAO) scale from multipoles of the redshift-space correlation function as measured by a low angular resolution 21cm IM survey of this kind. Our approach is to construct an analytic model of the multipoles of the correlation function and their covariance matrix that includes foreground contamination and beam resolution effects, which we then use to generate an ensemble of mock data vectors from which we attempt to recover the BAO scale.
In line with previous studies, we find that recovery of the transverse BAO scale $\alpha_{\perp}$ is hampered by the strong smoothing effect of the instrumental beam with increasing redshift, while the radial scale $\alpha_\parallel$ is much more robust. The multipole formalism naturally incorporates transverse information when it is available however, and so there is no need to perform a radial-only analysis. In particular, the quadrupole of the correlation function preserves a distinctive BAO `bump' feature even for large smoothing scales. We also investigate the robustness of BAO scale recovery to beam model accuracy, severity of the foreground removal cuts, and accuracy of the covariance matrix model, finding in all cases that the radial BAO scale can be recovered in an accurate, unbiased manner.
\end{abstract}




\section{Introduction}
As a mode of tracing the Universe's large-scale structure, neutral hydrogen (HI) intensity mapping (IM) will likely be unmatched in its capacity to survey the matter distribution of very large volumes efficiently. Rather than restricting attention to individual sources, the IM technique produces a 3D image of the total intensity from the combination of all objects that are found within each resolution element, or voxel. The HI line makes an excellent target for this method in cosmology due to its ubiquity; being found within galaxies as shielded clumps at late times. The hyperfine spin-flip transition of neutral hydrogen that occurs at $\lambda \simeq 21.1$cm allows distance measurements to made be with high fidelity, since they are deduced directly from the line's redshift, with accuracy only dependent on the frequency resolution of the observing radio telescope. Under the assumption that HI traces the underlying cosmological matter distribution with some associated bias, this method makes it possible to survey large swathes of the matter distribution out to very high redshift in a comparatively short observing time \citep{Bharadwaj2000, Battye2004, McQuinn2005, Mao2008, Chang2007, Wyithe2007, Loeb2008, Pritchard2008, Peterson2009, Bagla2009, Seo2009, Ansari2011}. During epochs when the neutral hydrogen abundance/ionisation fraction is evolving rapidly, 21cm IM can also be used to probe the various astrophysical processes that contribute to ionising the inter-galactic medium \citep{1997ApJ...475..429M, Barkana:2004zy, Barkana:2004vb, Mesinger:2007pd, Pritchard2008, 2014ApJ...782...66P}.

Different observing strategies can be deployed to measure 21cm intensity maps at various epochs, each with their own set of advantages and drawbacks. Interferometric experiments typically allow smaller angular scales to be accessed, with a maximum resolution set by the largest separation between dishes in the array. Often constructed as dense arrays, and used in a tracking or drift-scan mode, interferometers are advantageous in terms of their instrumental stability, but sample only a subset of the available angular Fourier modes, and tend to suffer from strong chromatic effects that can mix bright foreground contamination into otherwise signal-dominated modes. Alternatively, observations can be carried out in autocorrelation or `single-dish' mode, where each receiver in the array independently measures the total power signal at each pointing. Autocorrelation observations have been proposed as a way of accessing the largest cosmological scales, which are typically resolved out by interferometers, as well as for improving the sensitivity and survey speed of sparse arrays \citep{Battye2012, Bull2015, Santos2017}. Their angular resolution is limited by the dish size, which for modern multi-dish arrays with $\sim 15$m dishes translates to an angular resolution of order a degree at $z \sim 1$.  While their response is less chromatic than for an interferometer, autocorrelation instruments suffer from correlated ($1/f$) noise, and so must typically scan rapidly across the sky in order to avoid striping artifacts. This results in reduced stability of the system, leading to additional time-dependent systematic effects that must be filtered out of the data before maps are constructed.

While a wide variety of 21cm IM surveys are currently either underway or in the advanced stages of planning and construction, a definitive detection of the cosmological 21cm signal is yet to have been achieved at either high or low redshift, with the exception of detections in cross-correlation with optical galaxy surveys by GBT \citep{Wolz2021} and Parkes \citep{Anderson2018}. The reason for this is largely due to the difficulty of calibrating and processing 21cm data with sufficient fidelity; observations are dominated by foreground contamination from our Galaxy and extragalactic sources that are in excess of 3 orders of magnitude brighter than the expected cosmological signal \citep{Oh2003, Santos2004}, necessitating extremely precise instrumental calibration that strongly suppresses the leakage of foreground power into signal-dominated modes. It is possible to make significant headway in the removal of foregrounds, as they are expected to be smooth functions of frequency that can in principle be filtered out with only a small loss in the recovered cosmological signal \citep{Wang2005, Liu2009, Liu2011, Petrovic2010, Wolz2013, Shaw2014, Alonso2014, Wolz2015, Cunnington2019, Soares2021}. This is complicated by the chromaticity of the instrumental beam effect however, which is in general a non-trivial 2D sensitivity function that changes with frequency and receiver geometry. In the single-dish configuration, the beam function is convolved with the observed intensity field and produces a frequency-dependent smoothing effect that not only dampens features at or below the scale of the beam size, but also modulates the foregrounds, resulting in foreground power being scattered to Fourier modes at higher wavenumbers \citep{Santos2004, Jelic2008,  Chapman2012, Villaescusa-Navarro2017, Asad2019, Matshawule2020}. In interferometry, the chromatic beam instead acts as a window function on the intensity field, and has significant interaction with foreground removal algorithms \citep{Liu2014, Choudhuri2020, Hothi2020}. Nevertheless, advances in calibration and signal filtering are gradually improving measurements to the point that positive detections of the cosmological 21cm signal are anticipated in the coming years without the need for cross-correlation \citep{McKinley2018, Wang2020,Thyagarajan2020}.

In this paper, we consider the effects of foreground contamination and beam smoothing on the recovery of one of the key cosmological distance indicators -- the Baryon Acoustic Oscillation (BAO) scale -- in autocorrelation intensity maps of the kind that will be measured by the MeerKAT radio array. The BAO are acoustic waves in the pre-recombination photon-baryon plasma driven by gravitational interaction with dark matter and its own radiation pressure. Waves at the scale of the sound horizon froze into the matter distribution at the time of recombination, leaving a strong imprint that we are able to detect in the 2pt correlation function, the feature appearing as a local maximum at approximately 100 $h^{-1}$Mpc. Measurements of the Cosmic Microwave Background constrain the sound horizon scale, allowing the BAO feature to be used as a cosmological `standard ruler' \citep{Eisenstein1998} that can be used to derive constraints on the Hubble parameter, the angular diameter distance, and also the growth rate through the effects of redshift-space distortions. The BAO scale is well within the linear regime and stands out from the background continuum in the correlation function, and so it is difficult to confuse with systematic effects \citep{Eisenstein2007, Crocce2008, Padmanabhan2009}. This robustness to systematics is what makes BAO an optimal target for initial applications of the 21cm IM method as the technique advances in efficacy. 

The BAO scale has been measured variously in galaxy clustering surveys \citep{Cole2005, Eisenstein2005, Blake2007, Anderson2013, Beutler2017, Alam2016, Slepian2016}, the Ly-$\alpha$ forest \citep{Font-Ribera2014, Delubac2014}, and voids \citep{Liang2015, Kitaura2015}. The precision of these measurements can often be further boosted by using algorithms that reconstruct the linear BAO peak using non-linear density field information \citep{Eisenstein2006, Padmanabhan2012, Nikakhtar2021}. 21cm IM surveys have the potential to effectively `complete' the task of BAO measurement, as they can in principle measure the BAO scale over the full redshift range out to the Epoch of Reionization ($z \gtrsim 6$), and over almost the full sky \citep{Bull2015, Bull2015a,Obuljen2016,Bandura2019}.

In the coming decade, the Square Kilometre Array\footnote{\url{https://www.skatelescope.org/}} (SKAO) will be able to measure the 21cm cosmological signal at multiple stages of cosmic history using the autocorrelation technique. The SKAO's Mid telescope is a multi-dish radio array that will soon begin construction in the Karoo desert of South Africa. Part of the Mid telescope will comprise of MeerKAT, a 64-dish array that is already operational on the SKAO site \citep{Santos2017}. Combined with a low-frequency array sited in Australia, SKAO will eventually have the capacity to make very high resolution maps of the 21cm line from $z \simeq 0$ all the way out to $z \simeq 27$, well past the Epoch of Reionization (EoR) and into the Cosmic Dawn, where it has the potential to spatially resolve bubble structures around the very first stars and galaxies. Though the instrument will have unprecedented raw sensitivity, the data analysis for this survey represents an exceptional calibration challenge \citep{Wang2020}.

In this paper we seek to understand how instrumental beam smoothing and foreground filtering will affect the observed 2D correlation function and its covariance in the case of the MeerKLASS survey, a 4,000 deg$^2$, 4,000 hour precursor survey in the L-band ($900-1670$ MHz , $0 \leq z \leq 0.57$) with MeerKAT \citep{Santos2017}. In this work, we will consider a single redshift band centred at $z = 0.39$ that avoids surrounding RFI-contaminated regions. A second band at lower redshift has also been observed by MeerKAT \citep{Wang2020}, but we ignore it here as it covers too small a volume. In particular, we wish to assess how recovery of the BAO feature might proceed under various analysis assumptions, with the goal of identifying a viable strategy for a first detection with this instrument. Instead of performing a computationally-expensive analysis using simulations of the full survey, we use a partially-analytic approach in which the analytic models for the signal and covariance are used to generate noisy realisations of the observed 21cm correlation function under different analysis assumptions. We then perform a simulated analysis on these mock data using a combination of least-squares model fitting and Monte-Carlo Markov Chain (MCMC) analysis.

The recovery of the BAO feature in an SKAO-like 21cm autocorrelation survey has been studied previously. Most analyses have taken a purely Fourier-space approach \citep[e.g.][]{Bull2015, Soares2021}, in which models for the 2D redshift-space power spectrum can be fitted directly to the data. While this is a powerful approach, careful handling of systematic effects and survey window functions is required in order to avoid mode-coupling and subsequent leakage/scattering of foreground power outside of nominally foreground-contaminated regions.
This adds extra complexity to the analysis. Instead, we focus on the redshift-space correlation function as a slightly more conservative approach to obtaining an initial detection.

This paper is organized as follows. In Section~\ref{sec:corrfn} we describe our modelling of the multipoles of the 21cm correlation function in the presence of realistic instrumental beam effects and a foreground cut. We also derive an analytic covariance model for the multipoles in the presence of these effects. In Section~\ref{sec:methods} we describe our analysis methods, including our specific assumptions about the MeerKAT configuration and the function fitting and BAO recovery techniques that we have used. In Section~\ref{sec:results} we present our results for the correlation function and covariance matrix in realistic scenarios for MeerKAT, and the results of fits aimed at identifying the best analysis choices for the MeerKLASS survey.  Section~\ref{sec:conclusions} contains our conclusions.

\section{The 21cm correlation function and its covariance}
\label{sec:corrfn}

\begin{figure}
    \centering
    \includegraphics[width=\columnwidth]{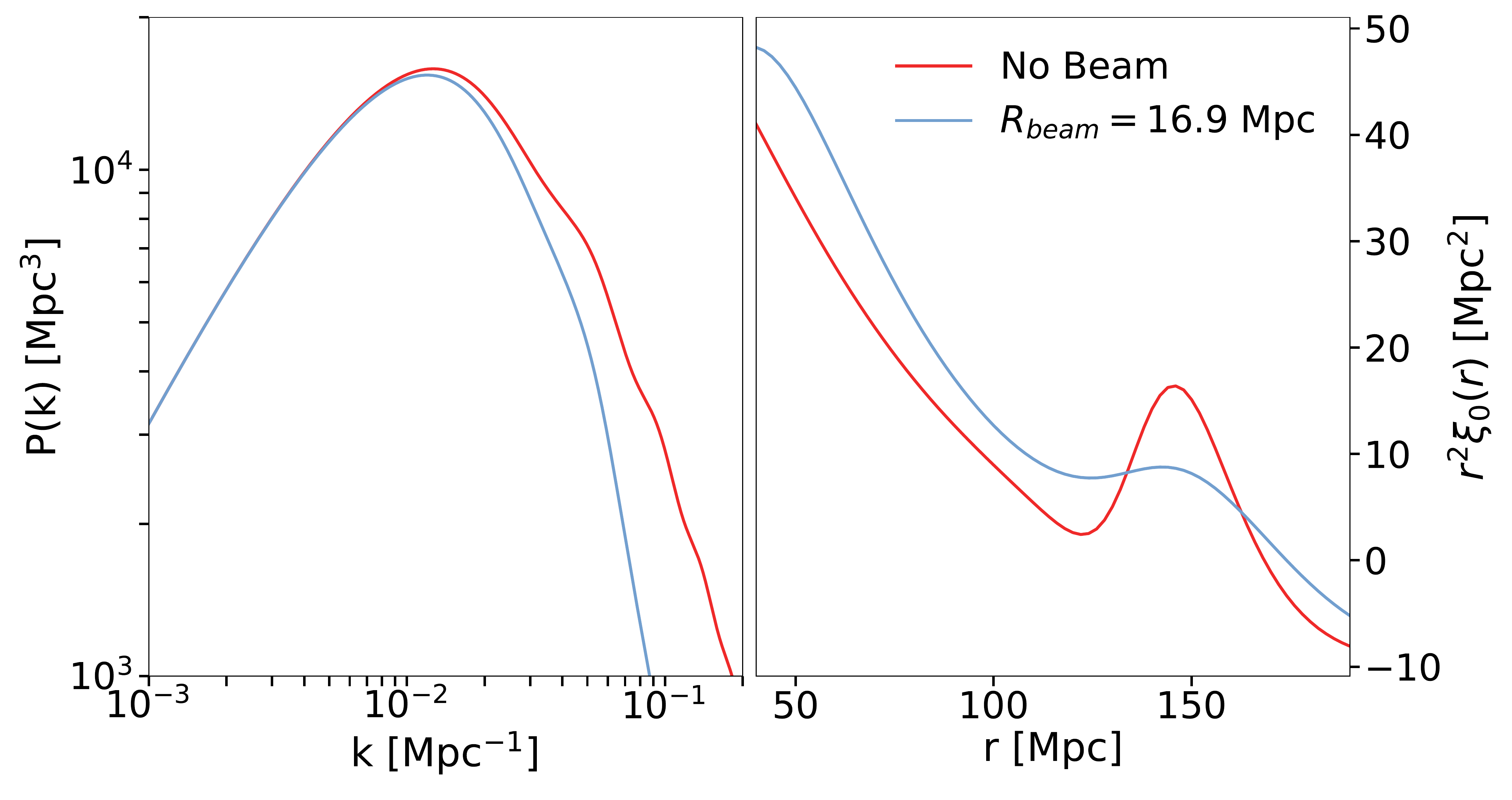}
    \caption{The linear power spectrum and resulting monopole of the 2pt correlation function, $\xi_0(r)$, shown with and without the effects of the MeerKAT beam response at $z=0.3915$ ($R_{\text{beam}} = 16.9$ Mpc). The BAO wiggles are significantly damped, and the corresponding BAO peak feature in the correlation function is smoothed.}
    \label{pkxir}
\end{figure}

The 21cm correlation function was studied by \cite{Villaescusa-Navarro2017}, who showed that the transverse smoothing effect due to the instrumental beam effectively washes out the BAO feature in the monopole of the correlation function at all but the lowest redshifts for an instrument like MeerKAT, making it impossible to disentangle from the smooth continuum of the correlation function. Figure \ref{pkxir} shows this effect for a MeerKAT-like beam response on the linear power spectrum and the 2pt correlation function at $z=0.3915$. Instead, they advocate for a line-of-sight only analysis, averaging out the transverse modes in Fourier space to form a 1D ($k_\parallel$-only) power spectrum. While this necessarily destroys any residual information about the BAO scale in the transverse direction, the BAO feature remains distinctive in the resulting 1D power spectrum. We adopt an alternative approach that strictly only uses the redshift-space correlation function, decomposing it into multipoles in an attempt to preserve as much information about the BAO scale as possible.
While the transverse modes are heavily smoothed by the beam response, they still contain some useful information, which it is possible to extract with appropriate beam modelling. Importantly, we derive an analytic model for the covariance matrix of the monopole and quadrupole of the redshift-space 21cm correlation function in the presence of both realistic beam smoothing and foreground removal systematics, allowing us to optimise the recovery of information.

In this section we derive analytic expressions for the redshift-space 21cm correlation function, its multipoles, and their covariance, including the effects of redshift-space distortions (RSD), the instrumental beam, and a foreground cut on line-of-sight ($k_\parallel$) modes. This extends well-known results for galaxy surveys that include the effects of RSDs only. Despite the added complications, we find that the 21cm correlation function can be calculated in a relatively inexpensive way via this multipole expansion, and present an implementation (including public code) that uses {\tt FFTLog} to speed up the calculation.

\subsection{The 2D correlation function}

We consider a scenario in which the anisotropic effects of the instrumental beam and foreground cut respect azimuthal symmetry around the line of sight direction, so that we can work in a 2D (transverse and radial) coordinate system, making use of the flat-sky, distant observer approximation. Our scale of interest, the BAO scale, falls at approximately 1 degree, and corrections to this approximation are expected to be at the sub-0.1\% level in this redshift range \citep[see, e.g.][]{Matthewson:2020rdt}. Under these conditions, the 2D correlation function as a function of components of the comoving separation $(r_\perp, r_\parallel)$ is related to the 2D power spectrum as a function of wavenumbers $(k_\perp, k_\parallel)$ by a Fourier transform. We take an isotropic model of the power spectrum $P(k)$, and denote the entire anisotropic modulation of the power spectrum, i.e. the effects of RSD, the beam, and foreground cut, as a function $F(k,\nu)$, such that
\begin{equation}
P_{\rm obs}(k, \nu) = F(k, \nu) P(k),
\end{equation}
where $\nu$ is a direction cosine defined below.
Note that we will define $F$ to include all of the tracer-dependent contributions to the observed signal, which in the case of 21cm IM will include a HI bias term and an overall brightness temperature.
Explicit models for the anisotropic modulation are given in Sect.~\ref{sec:methods}. We define the telescope pointing, or line-of-sight, direction to be $\mathbf{\hat n}$, and the real-space separation unit vector pointing radially outwards from the centre of the survey volume to be $\mathbf{\hat r}$. For the direction cosine between the telescope pointing and the separation vector we use the symbol $\mu \equiv \mathbf{\hat r} \cdot \mathbf{\hat n}$. The harmonic-space unit wave vector, which is the Fourier conjugate to $\mathbf{\hat r}$, is denoted by $\mathbf{\hat k}$, and the direction cosine between the telescope pointing and the $k$-mode is $\nu \equiv \mathbf{\hat k} \cdot \mathbf{\hat n}$. To be clear, $\mu$ is the direction cosine between the telescope pointing and a given real-space separation vector, and $\nu$ is the direction cosine between the telescope pointing and a given wave-vector. In this notation, the anisotropic correlation function is given by the Fourier transform of the total power spectrum, 
\begin{equation}
    \xi(r,\mu) = \frac{1}{(2\pi)^3} \int d^3k F(k,\nu) P(k) e^{i \mathbf{k} \cdot \mathbf{r}}. \label{eq:corrfn2d}
\end{equation}
Next, we substitute in the plane wave expansion,
\begin{equation}
    e^{i\mathbf{k} \cdot \mathbf{r}} = \sum_{\ell=0}^{\infty} i^{\ell} (2\ell+1) \mathcal{P}_{\ell}(\mathbf{\hat k} \cdot \mathbf{\hat r}) j_{\ell}(kr),
\end{equation}
where $\mathcal{P}_{\ell}(x)$ and $j_{\ell}(x)$ are the Legendre polynomials and spherical Bessel functions of order $\ell$ respectively, and also carry out a multipole expansion of the anisotropic modulation, $F(k,\nu)$. A general multipole expansion decomposes an angular function into radially-dependent coefficients of the Legendre polynomials,
\begin{equation}
    F(k, \nu) = \sum^{\infty}_{\ell=0} c^{(1)}_{\ell}(k) \mathcal{P}_{\ell} (\nu).
\end{equation}
The expansion coefficients $c^{(n)}_{\ell}(k)$ are determined using the orthogonality of the Legendre polynomials, where we introduce the notation
\begin{equation}
    c^{(n)}_{\ell}(k) = \frac{2\ell+1}{2}\int_{-1}^1 d\nu\, \mathcal{P}_{\ell}(\nu) [F(k,\nu)]^n,
\end{equation}
which will become useful when we consider the covariance calculation. The complete form of $F(k,\nu)$ is given in Eq. \ref{F-eqn}.
With these expansions in hand, the Fourier transform in Eq.~\ref{eq:corrfn2d} can now be expressed as
\begin{dmath}
    \xi(r,\mu) = \frac{1}{(2\pi)^3}  \int d^3k\, P(k)   \Bigg[ \sum_{\ell=0}^{\infty} i^{\ell} (2\ell+1) \mathcal{P}_{\ell}(\mathbf{\hat k} \cdot \mathbf{\hat r})  j_{\ell}(kr) \\ 
    ~~~~~~~~~~~~~~~~~~~~~~~~~~~~ \times \sum_{\ell^\prime = 0}^\infty c^{(1)}_{\ell^\prime}(k) \mathcal{P}_{\ell^\prime}(\nu)  \Bigg].
\end{dmath}
The angular integral is over a product of Legendre polynomials as a function of angles with respect to $\mathbf{\hat k}$. This can be expanded using the addition theorem of spherical harmonics,
\begin{dmath}
    \int d^2\Omega_k \mathcal{P}_{\ell}(\mathbf{\hat k} \cdot \mathbf{\hat r}) \mathcal{P}_{\ell^\prime}(\mathbf{\hat k} \cdot \mathbf{\hat n}) = \left(\frac{4\pi}{2\ell+1}\right)^2 \int d^2\Omega_{\mathbf{k}} \sum_{m=-\ell}^{\ell} Y_{\ell m}(\mathbf{\hat k}) Y^*_{\ell m}(\mathbf{\hat r})  \sum_{n=-\ell'}^{\ell'} Y^*_{\ell' n}(\mathbf{\hat k}) Y_{\ell' n}(\mathbf{\hat n}), \nonumber
\end{dmath}
where $Y_{\ell m}(x)$ the spherical harmonic of order $(\ell, m)$. As a consequence of the orthogonality of the spherical harmonics under integration, only terms satisfying $m=n$ are non-zero. Evaluating this integral and re-applying the addition theorem, we obtain
\begin{dmath}
    \int d^2\Omega_k \mathcal{P}_{\ell}(\mathbf{\hat k} \cdot \mathbf{\hat r}) \mathcal{P}_{\ell}(\mathbf{\hat k} \cdot \mathbf{\hat n}) = \delta_{\ell\ell^\prime} \frac{4\pi}{2\ell+1}\mathcal{P}_{\ell}(\mathbf{\hat n} \cdot \mathbf{\hat r}).
\end{dmath}
The action of $\delta_{\ell\ell^\prime}$ allows terms from each multipole expansion to be collected under a single summation. For brevity, we next combine the radial part of the integral for Legendre mode $\ell$ into the quantity
\begin{equation}
    I_{\ell}(r) = \int_0^{\infty} dk k^2 c^{(1)}_{\ell}(k) P(k) j_{\ell}(kr).
\end{equation}
The resulting final expression for the 2D correlation function reads as its own multipole expansion,
\begin{equation} \label{eq:corrfn_multipoles}
    \xi(r,\mu) = \sum_{\ell=0}^{\infty} \mathcal{P}_{\ell}(\mu) \frac{i^{\ell}}{2\pi^2}I_{\ell}(r).
\end{equation}
We can immediately see the useful result that the multipole coefficients of this expression have a straightforward form,
\begin{equation}
    \xi_{\ell}(r) = \frac{i^{\ell}}{2\pi^2} I_{\ell}(r).
\end{equation}
In other words, to calculate a given multipole $\ell$ of the correlation function, only the Legendre coefficient of the power spectrum modulation $c^{(1)}_{\ell}(k)$ of the same order is required. In what follows, we use the expression above as a model for the monopole $(\ell = 0)$ and quadrupole $(\ell = 2)$ of the correlation function.

\subsection{The covariance of $\xi_{\ell}(r)$}

We additionally construct an analytic model of the covariance of the multipoles of the 21cm correlation function, under the assumption that the bins of the correlation function can be approximated as being Gaussian distributed. The advantage of an analytic model is that the covariance can readily be calculated for a range of different instrumental configurations, cosmologies etc. without recourse to suites of expensive large-scale structure simulations. The main drawback is that non-linear effects and non-Gaussianities are left unmodelled. Since we are focusing on the BAO feature at large scales, we expect an analytic covariance model to be sufficiently accurate for our purposes here, although a more rigorous confirmation of this expectation is left for future work.

We construct the covariance by considering moments of the binned 2D correlation function. We begin by considering the 3D correlation function, which is the expectation value of the product of the matter density contrast at two points with a comoving separation $\mathbf{r}$,
\begin{equation}
    \xi(\mathbf{r}) = \langle\delta(\mathbf{x}) \delta(\mathbf{x + r}) \rangle.
\end{equation}
Under the assumption that the underlying density field is traced by a discrete set of objects (e.g. galaxies), there is an additional Poisson noise contribution to the observed correlation function, which we model as an uncorrelated shot noise term,
\begin{equation}
\xi_{\text{obs}}(\mathbf{r}) \equiv \left\langle\left(\delta(\mathbf{x}) + \frac{1}{\bar n}\right)\left(\delta(\mathbf{x}+\mathbf{r}) + \frac{1}{\bar n}\right)\right\rangle,
\end{equation}
where $\bar n$ is the spatial average of the number density of the tracer objects. Since $\xi_{\text{obs}}(\mathbf{r})$ does not have zero mean in general, its covariance is 
\begin{equation}
C(\mathbf{r},\mathbf{r'}) = \langle \xi_{\text{obs}}(\mathbf{r}) \xi_{\text{obs}}(\mathbf{r'}) \rangle - \langle \xi_{\text{obs}}(\mathbf{r}) \rangle\langle \xi_{\text{obs}}(\mathbf{r'}) \rangle.
\end{equation}
In general, this expression can be decomposed into a series of terms involving 4-point and 2-point correlators involving convolutions of $\xi_{\text{obs}}(\mathbf{r})$ with itself (\citet{Tansella2018}). Assuming Gaussianity, we can apply a Wick rotation to simplify the 4-point terms, and then apply the convolution theorem to obtain
\begin{align}
     C(\mathbf{r,r^\prime}) = & \frac{1}{V(2\pi^3)}\nonumber\\ \times &\int_{V}d^3k\Bigg[ \bigg(\frac{1}{\bar n^2} + \frac{2}{\bar n} P_{\text{obs}}(k,\nu) +  P_{\text{obs}}^2(k,\nu)\bigg)\nonumber \\ &\times\bigg(  e^{i\mathbf{k}\cdot (\mathbf{r-r^\prime})} + e^{i\mathbf{k}\cdot (\mathbf{r+r^\prime})} \bigg)\Bigg], 
\end{align}
where $V$ is the survey volume within which the correlation function is evaluated. 
The three separate contributions to the covariance are clear in the first set of square brackets in this expression: the first term is a pure shot noise contribution, the second term is a noise-clustering cross-term, and the last term constitutes the pure clustering term. An identical expression can be found in the calculation used by the {\tt COFFE} code \citep{Tansella2018}, following earlier work on cross-correlation covariances \citep{Bonvin2015, Hall2016}; see also \citet{Smith2009, Grieb2016} for another consideration of the binned covariance matrix. This expression can be further extended to take into account the finite size of survey redshift bins; evaluating the covariance at the central redshift of the bin is sufficient for our purposes so we do not take into account the redshift bin width except for in our specification of the spatial volume.

To further simplify this expression and introduce the multipole expansion of the correlation function, we once again substitute the plane-wave expansion for the complex exponential terms. The covariance of multipoles $(\ell,\ell^\prime)$ of $\xi(\mathbf{r})$ can then be obtained by evaluating the multipoles of the 3D covariance $C(\mathbf{r,r^\prime})$ for comoving separations $(r, r^\prime)$,
\begin{equation}
    \text{C}_{\ell\ell^\prime}(r,r^\prime) = \frac{(2\ell+1)(2\ell^\prime+1)}{4}\int_{-1}^1 d\mu  \int_{-1}^1 d\mu^\prime \mathcal{P}_{\ell}(\mu) \mathcal{P}_{\ell^\prime}(\mu^\prime)C(\mathbf{r,r^\prime}). \nonumber
\end{equation}
After further simplifications that make use of the properties of Legendre polynomials and Bessel functions (see Appendix~\ref{sec:covderiv} for a derivation), the resulting expression is 
\begin{dmath} \label{eq:multipolecov}
    \text{C}_{\ell\ell^\prime}(r_i,r_j) =  \frac{i^{\ell-\ell^\prime}}{V\pi^2}
    \\ \times \Bigg(\frac{(2\ell+1)\pi}{2\bar n^2 L_p r^2} \delta_{ij} \delta_{\ell\ell^\prime} + \frac{2}{\bar n}A_{\ell\ell^\prime}(r_i,r_j) + B_{\ell\ell^\prime}(r_i,r_j) \Bigg),
\end{dmath}
where $L_p$ is the size of each side of the 3D voxels used to calculate the covariance (i.e. corresponding to the binning of the 3D correlation function), and $V$ is again the survey (redshift bin) volume. The functions $A$ and $B$ are defined by making use of the Wigner 3-j symbol $\mathcal{W}$,
\begin{align}
    A_{\ell\ell'}(r_i,r_j)& =  ~(2\ell+1)(2\ell'+1)\nonumber \\
    & \times \int_0^{\infty}dk k^2 P(k) j_{\ell}(kr_i) j_{\ell'}(r_j) \sum_{n} c^{(1)}_n(k) \bigg(\mathcal{W}^{\ell\ell'n}_{000}\bigg)^2 \nonumber
\end{align}
\begin{align}
    B_{\ell\ell^\prime}(r_i,r_j)& = ~(2\ell+1)(2\ell^\prime+1)  \nonumber \\
    & \times \int_0^{\infty}dk k^2 P^2(k) j_{\ell}(kr_i) j_{\ell^\prime}(r_j) \sum_{n}c^{(2)}_n(k) \bigg(\mathcal{W}^{\ell\ell^\prime n}_{000}\bigg)^2 \nonumber.
\end{align}
In the aforementioned {\tt COFFE} covariance calculation, the effects of RSDs are handled analytically, leading to a similar non-trivial multipole expansion of the covariance as shown above. Our implementation extends this to include additional anisotropic effects that are present in 21cm data, including the instrumental beam and a foreground cut. An important difference is that the multipole coefficients of these effects are functions of $k$ in general, rather than being constant as is the case for the RSDs, and so $A$ and $B$ now include the multipole coefficients $c^{(n)}_{\ell}$ inside the integrals.

In Section~\ref{sec:methods}, we will evaluate the multipole coefficients, and hence the correlation function and its covariance, for particular choices of instrumental beam model and foreground cut. Our computations use a fast method for evaluating the integral $I_\ell(r)$ based on {\tt FFTLog}, which we outline in Appendix~\ref{app:fftlog}.

\subsection{Noise contribution}
\label{sec:noise}

In the expressions above, we have included an uncorrelated shot noise contribution to the observed correlation function, which is the main source of noise in galaxy surveys. While a small shot noise contribution is also expected to be present in the 21cm signal, the dominant source of noise is instead expected to be thermal noise due to the overall temperature of the receiver system, modelled by the system temperature, $T_{\rm sys}$. Since this is also an uncorrelated random component with mean zero, we can include it in our model without any further changes to the expressions above, simply by writing its contribution to the variance as an effective number density. For an autocorrelation experiment, this can be derived from the radiometer equation to obtain
\begin{equation}
\frac{1}{n_{\rm IM}} =  (\Delta\tilde\nu\, S_{\text{area}}) (r^2r_{\nu}) \frac{\mathcal{I}}{\Delta\nu \,t_{\text{tot}}}  \bigg(\frac{T_{\text{sys}}}{T_b}\bigg)^2,
\label{eq:nim}
\end{equation}
where $\Delta \nu$ is the frequency bin width; $\Delta\tilde\nu = \Delta \nu / \nu_{\text{21cm}}$ is the dimensionless redshift bin width; $\mathcal{I} = N_{\rm dish}^{-1}$ is a dish multiplicity factor; $r$ is the comoving distance to the centre of the redshift bin; $r_{\nu} = c(1+z)^2 / H(z)$ is a redshift to distance conversion factor; $S_{\text{area}}$ is the area of the sky covered by the survey; $t_{\text{tot}}$ is the total integration time; $T_{\text{sys}}$ is the system temperature; and $T_b$ is the HI brightness temperature. The leading factors in parentheses correspond to the redshift bin volume in observed coordinates (first term) and the conversion to comoving units (second term).
A slightly different approach was taken in \citet{Bull2015}, where an {\it anisotropic} effective number density was constructed that also included the effect of the instrumental beam. It is important to note that this choice was made for convenience; in the Fisher matrix expressions used in \citet{Bull2015}, the beam effect could be attached to either the signal or noise power spectrum terms without any loss of generality. In this paper, we have consistently included the beam effect as part of the signal power spectrum model, and so the noise term is isotropic and scale-independent.

\section{Recovery of the BAO scale}
\label{sec:methods}
In this section we describe our methods for recovering the radial and transverse BAO scale from simulated (mock) measurements of the multipoles of the 21cm correlation function from a MeerKAT IM survey. We begin by defining a model of the 21cm power spectrum that includes an anisotropic `shift' parametrisation of the BAO feature, a realistic instrumental beam smoothing effect, redshift-space distortions, and the effects of a foreground cut. We describe the specific models we use for each of these anisotropic effects, followed by a set of phenomenological fitting models for de-trending the continuum of the correlation function and recovering the BAO feature using a simple model fitting procedure. Finally, we outline the parameters of a fiducial 21cm IM survey with MeerKAT, based on the proposed MeerKLASS survey specification \citep{Santos2017}.

In what follows, we use the {\tt CCL} cosmology library \citep{2019ApJS..242....2C} to calculate background quantities and the linear matter power spectrum in our fiducial cosmology, defined by $\Omega_m, \Omega_b, h, n_s,\sigma_8 $ = \{0.315, 0.049, 0.67, 0.96, 0.83\} obtained from \cite{Ade2014}.

\subsection{Shift parameterisation of the power spectrum}

We wish to construct a simple phenomenological model for the observed monopole and quadrupole of the 21cm correlation function that can be used to extract the radial and transverse BAO scales in an unbiased way. While in principle we could construct a detailed forward model of the data based on the analytic models from the previous section, this would be computationally intensive if used in a model-fitting procedure. By using a simpler phenomenological fitting model instead, where features such as the smooth continuum of the correlation function are fitted out using (e.g.) polynomials, we are able to obtain results much faster. This procedure is also closer to what is typically used to extract the BAO feature from galaxy surveys.

Our phenomenological model is based on the common strategy of parameterising deviations from a fiducial cosmological model. 
The BAO feature, or specifically the departure of the BAO scale from that found within the fiducial cosmology, may be parameterised by introducing a pair of `shift' parameters, $\alpha_{\perp}, \alpha_{\parallel}$. These parameters represent the departure from the fiducial values of the angular diameter distance $D_A(z)$ and expansion rate $H(z)$,
\begin{equation}
    \alpha_{\perp} = \frac{D_A(z)}{D_A^{\text{fiducial}}(z)}; ~~~~~~
    \alpha_{\parallel} = \frac{H(z)^{\text{fiducial}}}{H(z)}. \label{eq:shift}
\end{equation}
Following (e.g.) \citet{Blake2003, Bull2015}, we first decompose the isotropic linear matter power spectrum $P(k)$ into smooth and oscillatory parts, $P_{\text{smooth}}$ and $f_{\text{BAO}}$ respectively, 
\begin{equation}
P(k,k^\prime,z) = \bigg(1+Af_{\text{BAO}}(k^\prime,z)\bigg)P_{\text{smooth}}(k,z)
\end{equation}
where $A=1$ is the amplitude of the BAO feature, and $k^\prime$ denotes the wavenumber after an anisotropic shift has been applied,
\begin{equation}
    k^\prime = \sqrt{(\alpha_\perp k_\perp)^2 + (\alpha_\parallel k_\parallel)^2} = \sqrt{\big(\alpha_{\perp}k\big)^2(1-\mu^2) + \big(\alpha_{\parallel}k\mu\big)^2}.
\end{equation}
Note that we only allow the anisotropic shift to affect the BAO feature. This choice ensures that only the recovered BAO feature imparts information about the shift parameters when we perform the model fits; the smooth power spectrum is assumed constant. In reality, deviations from the fiducial cosmology also result in an anisotropic shift in the broadband shape of the power spectrum, but extracting this information requires substantially more careful modelling however, which we forego here.

To split the power spectrum into smooth and oscillatory parts, we take a cubic spline over the linear power spectrum in log-log space, using only the points outside of the BAO region that we define as $0.017 < k < 0.45$ Mpc$^{-1}$. In non-logarithmic space, this spline represents the smooth part of the power spectrum, $P_{\text{smooth}}(k,z)$. The oscillatory part, $f_{\text{BAO}}(k^\prime,z)$, is then found by dividing the total power spectrum by the smooth part. 

The frequency of the oscillations in the harmonic space $f_{\text{BAO}}$ function effectively determines the separation at which the BAO feature will appear in the correlation function multipoles (\citet{Eisenstein2006}). An increase in either $\alpha_{\perp}$ or $\alpha_{\parallel}$ equates to the acoustic peak appearing at a greater separation in the correlation function, while any shift $\alpha_\parallel \neq \alpha_\perp$ introduces anisotropy into the correlation function.

\subsection{Anisotropic model of the observed power spectrum}

In the previous section, we constructed a model of the linear matter power spectrum with a BAO feature that shifts/stretches anisotropically depending on deviations from the fiducial background cosmology, according to Eq.~\ref{eq:shift}. For the purposes of this paper, this represents the full cosmological information content that we hope to be able to extract from the 21cm correlation function. In this section, we will incorporate a further set of observational effects that also contribute to the anisotropy of the observed power spectrum, and therefore of the 21cm correlation function, but which are in some sense `nuisance' effects that degrade our ability to recover the BAO scale.

Autocorrelation experiments observe the brightness temperature fluctuations of the redshifted 21cm line as a function of frequency and angle on the sky. By treating the neutral hydrogen as a linearly-biased tracer and converting HI mass density to brightness temperature, we can link the brightness temperature fluctuations to matter density fluctuations $\delta_M$, 
\begin{equation}
\delta T_b(\mathbf{k_{\perp}}, k_\parallel, z) = \bar T_{\text{HI}}(z)\, b_{\text{HI}}(z)\,\delta_{M}(\mathbf{k_{\perp}}, k_\parallel, z)
\end{equation}
with the mean brightness temperature given by 
\begin{equation}
    \bar T_{\text{HI}}(z) \approx 180 h\,\Omega_{\text{HI}}(z)\, \frac{(1+z)^2}{H(z)/H_0} \text{mK},
\end{equation}
where $z$ refers to the mean redshift of the band under consideration, and $\Omega_{\text{HI}}(z)$ is the HI fractional density at redshift $z$,  \citep{2012arXiv1209.1041B, Hall:2012wd, Bull2015}.
Wavelength maps to observed redshift according to $\lambda = \lambda_{\rm 21cm}\,(1+z)$, where $\lambda_{\rm 21cm} = 0.211$~m. To convert observed redshift and angular position into comoving coordinates, we must also account for peculiar velocities, which distort the mapping between `real space' and `redshift space'. See \cite{Hall:2012wd} for a careful treatment of this mapping that includes all relevant effects to linear order. In this paper, we will include only the effects of peculiar velocities, via a redshift-space distortion term $P_{\rm RSD}$ that multiplies the power spectrum, and neglect relativistic and wide-angle corrections.

The process of observing the redshift-space 21cm brightness temperature fluctuation field with an autocorrelation experiment imposes additional anisotropic effects on the signal. First, what is observed is a convolution of the true sky brightness temperature distribution with an instrumental beam function. In harmonic space, this can be represented as the product of the Fourier-transformed, wavelength-dependent beam power pattern, $B(\mathbf{k_\perp}, \lambda)$, with the brightness temperature fluctuations,
\begin{equation}
\delta T_b^{\rm obs}(\mathbf{k_{\perp}}, k_\parallel, z) = B(\mathbf{k_{\perp}}, z)\, \delta T_b(\mathbf{k_{\perp}}, k_\parallel,z).
\end{equation}
Note that $\mathbf{k_\perp}$ denotes a 2D vector in the plane of the sky; in what follows we will assume axisymmetry, in which case $B(\mathbf{k_\perp}, \lambda) \to B(k_\perp, \lambda)$, where $k_\perp = |\mathbf{k_\perp}|$.

Instrumental noise is also introduced into the observed signal, which we discussed in Sect.~\ref{sec:noise}. We assume this to be homogeneous, uncorrelated white noise, which does not impart any additional anisotropy into the measured correlation function. Finally, foreground contamination imparts a strongly anisotropic signal in Fourier space that is several orders of magnitude brighter than the target cosmological signal. This must be filtered or subtracted out in order to recover the cosmological signal, but all current foreground removal methods do this at the expense of losing cosmological signal in the overlapping region of Fourier space. The filtered data are therefore modulated by an anisotropic effective Fourier-space window function $W_{\rm fg}$ that accounts for the signal lost by the foreground removal process. The foreground removal process will leave residual unfiltered foregrounds in the data. We make the simplifying assumption that these residuals are uncorrelated and noise-like, and so would expect them to average down. We do not include an additional residual noise term in our analysis however.


Putting all of these effects together, we arrive at the following explicit form for the anisotropic modulation of the isotropic cosmological power spectrum:
\begin{align}
    F(k,\mu,z; \alpha_{\perp},\alpha_{\parallel}) = ~& \bigg[1+Af_{\text{BAO}}(k,\mu; \alpha_{\perp},\alpha_{\parallel})\bigg]\nonumber \\ & \times P_{\text{RSD}}(\mu, z)\, B^2(k_\perp, z)\, W_{\rm fg}(k,\mu,z),
    \label{F-eqn}
\end{align}
where the observed power spectrum is
\begin{equation}
P_{\rm obs}(k, \mu, z; \alpha_\perp, \alpha_\parallel) = F(k,\mu,z; \alpha_{\perp},\alpha_{\parallel}) P_{\rm smooth}(k, z).
\end{equation}
In the following sections, we construct explicit models for each of the anisotropic factors.

Note that there are other observational and instrumental effects that may cause anisotropies in the power spectrum that we have not modelled here. The excision of RFI and the shape of the survey region introduce a complex window function that can induce additional anisotropic structure into the analysis, particularly by coupling Fourier modes together \citep{2019MNRAS.484.2866O}. Correlated ($1/f$) noise, its coupling to the scan pattern of the instrument, and the filtering schemes used to mitigate it could also potentially introduce power anisotropies \citep{Bigot-Sazy:2015jaa, Harper:2017gln, Li:2020bcr}, as could polarisation leakage \citep{2014MNRAS.444.3183A, 2016ApJ...833..289L, Cunnington:2020njn}. It is also possible for calibration errors, for example due to beam or calibration source model errors, to also introduce additional anisotropic structure \citep{Matshawule2020}. We defer an examination of the impact of these effects on the correlation function to later work.

\subsection{Instrumental beam models}

The angular size of the MeerKAT instrumental beam ranges from around $0.9 - 1.4$ degrees in the redshift range covered by the L-band, which translates to only a factor of a few smaller than the angular scale of the BAO feature at the corresponding redshifts. The beam width grows with wavelength approximately according to $\theta_{\rm beam} \sim \lambda / D_{\rm dish} \approx 0.9 (1+z)\, {\rm deg}$, while at low redshift the angular size of the BAO feature scales approximately as $\theta_{\rm BAO} \sim 150\, {\rm Mpc} / (c z / H_0) \approx (2.0 / z)\, {\rm deg}$. As such, we expect beam smoothing to have an important effect on the observed 21cm correlation function that worsens with increasing redshift. Previous works have mostly studied this effect in Fourier space, where it is clear that BAO wiggles at higher $k$ are lost/down-weighted due to beam attenuation, but lower-$k$ wiggles remain intact even at relatively high redshifts, allowing some cosmological distance information to be recovered despite the poor angular resolution.

The picture is more complicated for the correlation function, which is related to the power spectrum by a Fourier transform. The BAO wiggles, encoded by the function $f_{\rm BAO}(k)$, resemble a wave packet. The frequency of the wiggles within the wave packet sets the scale at which the BAO feature appears in the correlation function, while the width of the packet sets the effective width of the feature. Beam attenuation effectively shrinks the wave packet in Fourier space, which corresponds to a broadening or smoothing of the feature in the correlation function. Even if one or two wiggles remain in the attenuated power spectrum, the reduction in packet width can cause such a strong degree of smoothing that a BAO bump feature is no longer discernible from the continuum of the correlation function. This effectively `hides' any remaining distance information from the BAO feature from detection in the correlation function, even though it is technically still there.

Since it is clear from this discussion that the recovery of the BAO scale will depend sensitively on the degree of beam smoothing, we attempt to work with as realistic a beam model as possible. 
We use the {\tt katbeam} package \citep{Matshawule2020} to model the MeerKAT beam response as a function of frequency. This makes use of electromagnetic simulations and field observations to construct detail beam models for both the L and UHF band receivers in both polarisations. We use {\tt katbeam} to generate the electric field beam, $E_i(\theta)$, at the centre frequency of each redshift bin, for angles in the range $[0^{\circ}$, $5^{\circ}]$ from beam centre for the HH polarization. The beam is close to being cylindrically-symmetric, and we use a single beam model to represent both polarisations. The electric field values are related to the beam power pattern by
\begin{equation}
    B(\theta) = |E_i(\theta)|^2.
    \label{beam-ei}
\end{equation}
We convert $B(\theta)$ to a function of transverse separation $B(r_{\perp})$ at the target redshift by stretching the $\theta$ axis by a factor of $\frac{\pi}{180}\,r(z)$ where $r(z)$ is the comoving (transverse) distance to redshift $z$ evaluated by {\tt CCL}. 

Since we have assumed that the beam has cylindrical symmetry, we can generate the harmonic-space beam function via a Hankel transform,
\begin{equation}
    B(k_{\perp}) = \int_0^{\infty} dr_{\perp} r_{\perp} J_0(k_{\perp}r_{\perp}) B(r_{\perp}).
\end{equation}
The resulting function is normalised to 1 at its maximum, and we then calculate its Legendre multipole coefficients. Note that the fundamental width of the MeerKAT beam has additional complicated behaviour in the frequency direction \citep{Asad2019, Matshawule2020}; for example, the beam width has a rapid low-level oscillation with frequency (which may introduce extra spectral structure through interactions with the foregrounds for example). We take the {\tt katbeam} outputs to have satisfactorily accounted for such effects, and do not attempt to refine the model any further.

\begin{figure}
    \centering
    \includegraphics[scale=0.26]{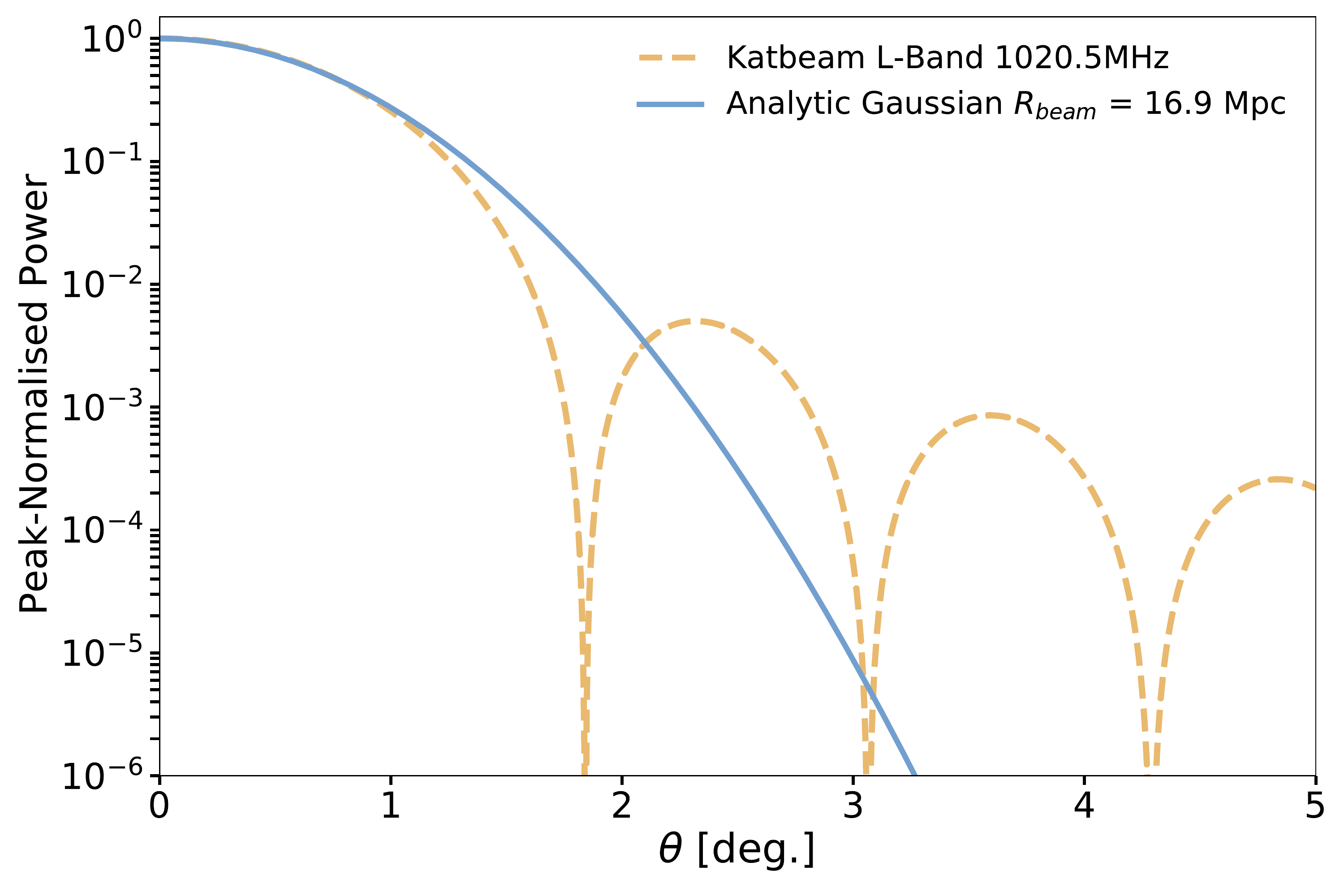}
    \caption{Comparison between the cylindrically-symmetrised {\tt katbeam} output (squared, see Eqn.~\ref{beam-ei}) for the MeerKAT beam model at $z=0.3915$, and a Gaussian beam model matched to its FWHM. The Gaussian beam model is a good approximation to the main lobe within 1 degree at this redshift, but does not capture the beam's side-lobes.} 
    \label{fig:beams}
\end{figure}

Since the beam smoothing effect enters the observed power spectrum expression as the square of the beam power pattern, we expect sidelobes to be greatly suppressed compared with the mainlobe. We therefore examine whether a much simpler beam model can be used that approximates only the mainlobe by a Gaussian with a FWHM matched to that of the true beam function. This approximation is advantageous since under a Hankel transform, a Gaussian transforms into another Gaussian, making this beam pattern particularly simple to work with. The analytic Hankel transform of a Gaussian real-space beam with standard deviation $R_{\text{beam}}$ is 
\begin{equation} \label{eq:gaussian_beam}
    B(k_{\perp}) = e^{-\frac{1}{2} k^2_{\perp}R_{\text{beam}}^2},
\end{equation}
and the multipole coefficients of its square, which we use in our correlation function analysis, are
\begin{equation}
    B^2_{\ell}(k) = \int_{-1}^1 \mathcal{P}_{\ell} (\nu)\, e^{-k^2 R_{\text{beam}}^2 (1-\nu^2) }\, d\nu,
\end{equation}
where $\mathcal{P}_{\ell}$ is the Legendre polynomial of degree $\ell$ and in this expression $\nu$ is the direction cosine between the line-of-sight direction and the Fourier wavevector.

\begin{figure}
    \centering
    \includegraphics[scale=0.26]{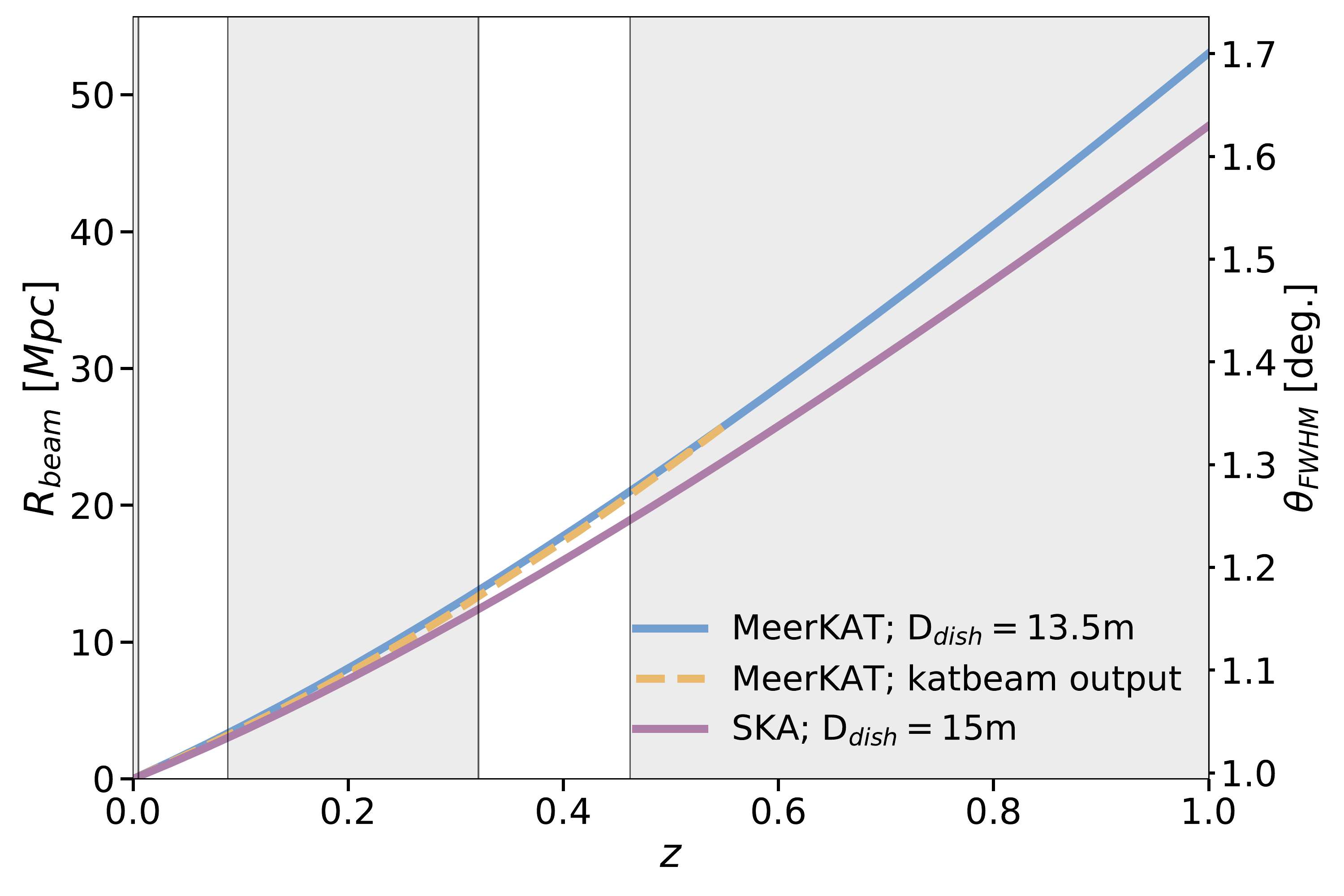}
    \caption{R$_{\text{beam}}$ values in the case of MeerKAT and the SKAO using $\theta_{\text{FWHM}} \approx {1.2\lambda / D_{\text{dish}}}$, and calculated from the outputs of the {\tt katbeam} package. Comoving distances from {\tt CCL}. White bands show redshift regions considered in this analysis. The region at lower redshift has not been studied further in this paper due to its small volume.}
    \label{fig:beamsize}
\end{figure}

We follow \cite{Villaescusa-Navarro2017} in defining the width of the effective Gaussian beam via
\begin{equation}
  R_{\text{beam}}  =  \frac{\theta_{\text{FWHM}}} {\sqrt{8 \ln 2}}  r(z).
\end{equation}
In order to determine the $R_{\text{beam}}$ values that match the width of the true MeerKAT beam, we construct a spline of the function $y = B(r_{\perp}) - 0.5$, find its root, and then multiply by 2 to determine $\theta_{\text{FWHM}}$.  Fig.~\ref{fig:beams} shows the {\tt katbeam} model at $z=0.3915$ as compared with the Gaussian beam model that is matched to its FWHM, while Fig.~\ref{fig:beamsize} shows how the resulting beam width varies with redshift.

Since the use of Gaussian beam models is relatively common in the literature, both the {\tt katbeam}-derived model and a Gaussian FWHM-matched model will be considered in the fitting analysis as an opportunity to better understand any interactions that the $R_{\text{beam}}$ may have with other fitting parameters.

\subsection{Redshift-space distortions and bias model}

Redshift-space distortions (RSDs) arise from the fact that we measure the position of sources in redshift rather than comoving distance. When observing a dense region along the line-of-sight, structures on the far side and near side will be subject to additional blue/redshifts respectively due to their infall velocity towards the overdensity. We use the linear RSD model according to \cite{Kaiser1987},
\begin{equation}
P_{\text{RSD}}(\mu,z) = \bigg(b_{\text{HI}}^2(z) + f(z)\mu^2 \bigg)^2,
\end{equation}
where $f(z)$ is the linear growth rate and $b_{\rm HI}$ is the linear bias factor that relates fluctuations in the HI density distribution to matter fluctuations. We have neglected the effects of non-linear velocities, e.g. the `Fingers of God' effect, which would contribute an additional suppression of power on small radial scales. For the HI bias, we use a simple fitting function derived from the bias model in \cite{Bull2015},
\begin{equation}
    b_{\text{HI}}(z) \approx \frac{b_{\text{HI},0}}{0.677} \left ( 0.667 + 0.178\,z + 0.0502\,z^2 \right),
\end{equation}
where $b_{\text{HI},0}$ is the amplitude of the HI bias function. We fix this factor to be equal to the denominator, i.e. $b_{\text{HI},0} = 0.677$. Note that the leading numerical factors do differ by a single digit.

\subsection{Foreground removal}

The impact of foreground cleaning methods on the recovery of the 21cm power spectrum is relatively well-studied for simulated data \citep[e.g.][]{Wolz2013, Alonso2014, 10.1093/mnras/stv2884, Cunnington2019, Carucci:2020enz, Cunnington:2020njn, 2020arXiv201015843M}. Since in this paper we do not construct full sky simulations, it is not possible to replicate the full effects of foreground cleaning algorithms on the recovered signal in detail. Within the scope of our analysis, we instead seek to model the basic effect of foreground removal, which is to effectively introduce a cut that removes the most foreground-contaminated Fourier modes.

For autocorrelation experiments, we do not expect to observe a `wedge' feature in Fourier space that affects interferometric observations \citep{thyag13, Thyagarajan2015, Seo2016}; instead, the foregrounds should remain confined to a region at low $k_\parallel$ with a width defined by chromatic effects due to gain errors and the instrumental beam \citep[e.g.][]{2013ApJ...763L..20M, Alonso2014, Cunnington2019}. We model this region as a Gaussian in $k_\parallel$ that suppresses modes below a cut-off $k_{\rm fg}$, 
\begin{equation}
    W_{\rm fg}(k,\mu) = 1 - \exp\left[-\frac{1}{2}\left(\frac{k_{\parallel}}{k_{\text{fg}}}\right)^2\right]
\end{equation}
where $k_{\parallel} = k\mu$ \citep{Bull2015, Soares2021}. This is broadly consistent with the signal suppression that would be expected from blind foreground removal methods that fit out smooth functions in the frequency direction. The smooth edges of the cut region have the advantage of reducing ringing in the Fourier transform when calculating the correlation function. This is equivalent to applying an apodisation to a Fourier-space foreground filter. We do not consider any dependence of the width of the region on $k_\perp$.

\subsection{Fitting the model to mock data}
\label{sec:fittingmethod}

Using a joint monopole and quadrupole model vector along with its covariance, we generate sets of Gaussian realisations that match the noise properties of the covariance. We then fit our model to these realisations and consider the fit distributions of $\alpha_{\perp},\alpha_{\parallel}$ that arise. 
The full fitting model for the multipoles of the correlation function is as follows
\begin{equation}
    \xi_{\ell,\text{fit}}(r) = D_{\ell}(r) + \frac{i^{\ell}}{2\pi^2}\int_0^{\infty} dk k^2 c^{(1)}_{\ell}(k) j_{\ell}(kr) 
\end{equation}
With the introduction of the $\alpha$-parameters, the power spectrum becomes a function of line-of-sight angle $\mu$ and hence must be included in the $c_{\ell}(k)$ calculation. The multipoles of the total power spectrum $c_{\ell}$ are as defined in section 2. The function $D_{\ell}(r)$ contains continuum fitting parameters. The monopole and quadrupole fitting parameters we use are comparable to \cite{Padmanabhan2012} for the monopole, and have inverted powers for the quadrupole:
\begin{equation} \label{eq:continuum}
    D_0(r) = a_0r + a_1 + \frac{a_2}{r} + \frac{a_3}{r^2}; ~~~~~~
    D_2(r) =  a_4 + a_5r + a_6r^2
\end{equation}
The fitting model then has 11 total parameters, which are
\begin{equation}
    \mathbf{\Theta} = \{ \alpha_{\perp}, \alpha_{\parallel}, A, R_{\text{beam}}, a_0, a_1, a_2, a_3, a_4, a_5, a_6\}
\end{equation}
Through testing we have found that using a range of separations $40-190$ Mpc for the monopole and $80-190$ Mpc for the quadrupole enables fitting to be carried out effectively, and that priors on each parameter determined through testing are also appropriate. We use prior ranges on both $\alpha_{\perp},\alpha_{\parallel}$ of \{0.7, 1.3\}, and consider fits at the edge of this region to be catastrophic failures, in the sense that they would be rejected if found in a real survey. Furthermore we fix the BAO amplitude parameter $A$ to its fiducial value of 1 and adopt a 5\% prior on the value of $R_{\text{beam}}$ in cases where its value is not fixed.

\begin{table}
\centering
\begin{tabular}{ |c|c| }
 \hline
 \hline
 $T_{\rm inst}$ & $\sim$16K \\
 Antennas & 64 \\
 Survey time & 4,000 hours \\
 Survey area & 4,000 deg$^2$ \\
  \multirow{2}{*}{Redshift bins} & [0.005, 0.088]  \\
  & [0.321, 0.462] \\
 Central redshifts & 0.0415, ~0.3915 \\
 \hline
\end{tabular}
\caption{Survey and instrumental specifications for a single-dish MeerKAT survey, similar to MeerKLASS.} \label{mkspecs}
\end{table}

We use the {\tt SciPy} routine {\tt curve\_fit} over sets of noisy realisations to test the recovery of $\alpha_{\perp},\alpha_{\parallel}$ under different observational effects and beam assumptions. We test the inclusion of each systematic in turn as well as their full combination. We also test how the total integration time should affect the noise in the fits, the effect of making use of the Gaussian approximation for the beam function when the data is convolved with the actual MeerKAT beam, and the impact of fitting with a sub-optimal covariance. We also make use of a likelihood method. Assuming a Gaussian distribution for each point in the joint-correlation function vector, the log-likelihood for a vector of measurements $\boldsymbol{\xi}$ we take to be
\begin{dmath} \label{eq:likelihood}
\mathcal{P} = -\frac{1}{2}(\boldsymbol{\xi} - \boldsymbol{\xi}_{\text{mdl}})^T\mathbf{C}^{-1} (\boldsymbol{\xi} - \boldsymbol{\xi}_{\text{mdl}}) +\frac{1}{4}\textmd{Tr}\big(\log \mathbf{C}\big) + \textrm{const.},
\end{dmath}
where $\mathbf{C}$ is the covariance matrix.

\subsection{Fiducial MeerKAT survey specification}

We adopt a similar survey specification to the MeerKLASS survey \citep{Santos2017}, assuming a 4,000 deg$^2$ sky area using MeerKAT 64 dual-polarisation receivers in the L-band operating in autocorrelation mode (Table~\ref{mkspecs}). The instrumental temperature in this band is $T_{\rm inst} \approx 16$~K. To calculate the system temperature, we include a mean sky temperature contribution of the form
\begin{equation}
T_{\rm sky} = 60\,(300\text{MHz} / \nu)^{2.55}\, {\rm K}.
\end{equation}
We assume an integration time per pointing of approximately 1.85 hours,
which corresponds to $n_{\rm IM} = 3.74 \times 10^{-3}$ Mpc$^{-3}$ (see Eq.~\ref{eq:nim}). Note that this is the effective integration time following duty cycle losses, including data lost to RFI flagging and noise diode fires that are used for calibration, which means that the actual observing time required to achieve this noise level is likely to be a factor of $\sim 2$ times longer.

Recent observations have also shown substantial segments of the MeerKAT L-band to be heavily polluted by RFI. We adopt the same frequency ranges as the analysis in \cite{Wang2020}, which conservatively avoids these regions of the band, resulting in two sub-bands, $971-1075$ MHz and $1305 - 1504$ MHz, where RFI is minimal. These are shown as white regions in Fig.~\ref{fig:beamsize}. For our analysis in the rest of the paper, we consider only the lower-frequency band, centred on $z = 0.3915$; the higher-frequency band covers a comoving volume of only $\sim (300\,{\rm Mpc})^3$ for a MeerKLASS-like survey area, making it highly sample variance-limited.

\section{Results}
\label{sec:results}
In this section we describe the effects of instrumental beams and a foreground cut on the correlation function, and present an analytic calculation of the covariance of the multipoles of the correlation function in the presence of these effects. We then demonstrate how they affect our ability to recover the radial and transverse BAO scales by performing model fits to large numbers of Gaussian random realisations of the binned correlation function multipoles with MeerKAT-like noise and beam specifications.

\subsection{The 2D correlation function}
In this section, we analyse how various anisotropic effects affect the 2D (redshift-space) 21cm correlation function. 
In Fig.~\ref{2dcorrs}, we plot the 2D correlation function calculated using Eq.~\ref{eq:corrfn_multipoles} after including each anisotropic effect in turn, beginning with the isotropic cosmology-only case, and then adding RSDs, beam smoothing, and a foreground cut respectively. To plot the correlation function, we sum multipoles up to $\ell = 25$, which is enough to suppress most artifacts that would arise if a smaller number of terms was used. For our calculation, we assume a MeerKAT-like configuration for a redshift bin centred at $z = 0.3915$, and do not include a noise contribution. Note that Fig.~\ref{2dcorrs} shows a smooth representation of $\xi(r_\perp, r_\parallel)$, and has not yet been binned in separation.

For clarity, Fig.~\ref{2dcorrs} shows the correlation function multiplied by the separation $r^2$ in order to enhance the visibility of the various features. The BAO feature is visible as an isotropic ring in the base cosmology case (first panel), and there is also an increase in correlation towards smaller separations, as expected. Once RSDs are added (second panel), the correlation function becomes strongly anisotropic; the BAO feature remains clearly visible for all angles with respect to the line of sight, but is most clearly defined in the purely radial direction ($r_\perp \approx 0$), where the underlying continuum has been suppressed.

When the beam response is added (third panel), the BAO feature is very clearly smoothed out in the purely transverse direction ($r_\parallel \approx 0$), and for a spread of angles around it. It has comparable sharpness to the no-beam case in the purely radial direction however. Note that some ray-like artifacts are visible at small separations in this panel; this is an artifact of the multipole expansion, and is increasingly strongly suppressed as more multipoles are included in the sum.

In the last panel, the addition of a foreground cut at $k_{\text{fg}} = 0.01$ Mpc$^{-1}$ pulls the correlation function down to strongly negative values in the radial direction, erasing the BAO feature and much of the continuum in a band of width $\Delta r_\perp \approx 50$ Mpc around $r_\perp = 0$. The BAO feature therefore only remains clearly visible at intermediate angles from the line of sight.

\subsection{Multipole covariance matrix}

\begin{figure}
    {\centering
	\hspace{-1em} 
	\includegraphics[scale=0.53]{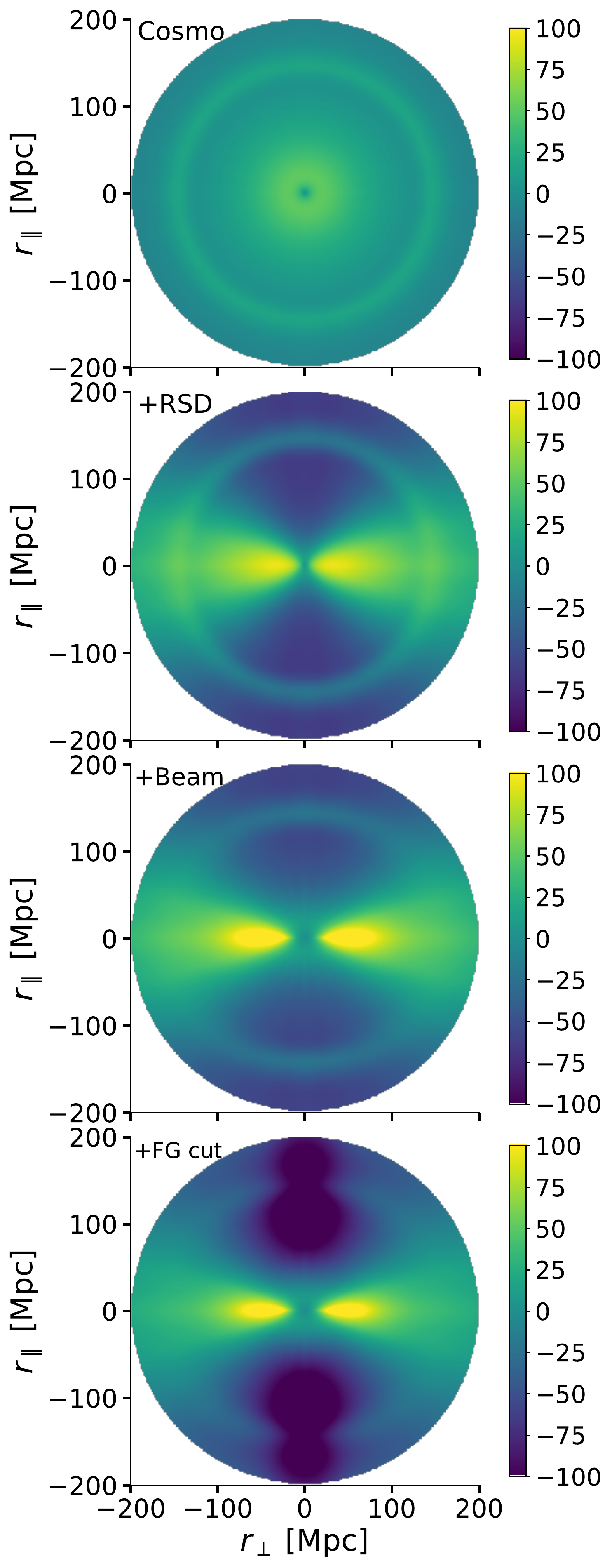}
	}
	\caption{Redshift-space correlation function models, $\xi^{\text{2D}}$, plotted as $r^2\xi^{\text{2D}}(r_\perp, r_\parallel)$ (in units of Mpc$^2$), as a series of anisotropic effects are cumulatively added. From top to bottom: isotropic cosmology-only case; Kaiser RSD term added (no Fingers of God); MeerKAT-like Gaussian beam added with $R_{\text{beam}} = 16.9$ Mpc; foreground cut at $k_{\parallel,\text{fg}} = 0.01$ Mpc$^{-1}$ added. Substantial anisotropic smoothing of the BAO feature is visible on addition of the beam response.}
	\label{2dcorrs}
\end{figure}

\begin{figure}
	{\centering
	\includegraphics[scale=0.53]{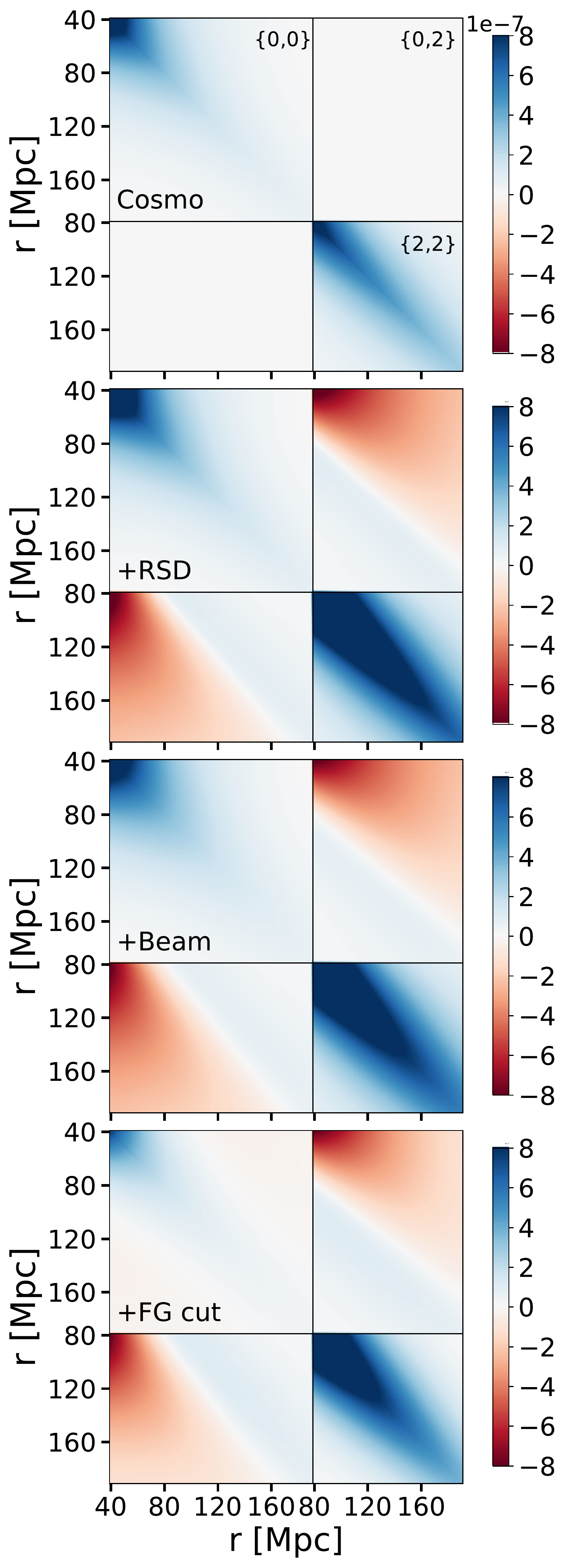}
	\hspace{-15em}}
	\caption{Block covariance plots at separations 40--190 Mpc for the monopole and 80--190 Mpc for the quadrupole as a series of anisotropic effects are cumulatively added to the model. The covariance shown here is dimensionless. From top to bottom: base cosmology only; with RSD added; with MeerKAT-like Gaussian beam added; with a foreground cut at $k = 0.01$ Mpc$^{-1}$ added.}
	\label{covs}
\end{figure}

Next, we study the effect of introducing the same anisotropic effects as in Fig.~\ref{2dcorrs} on the covariance matrix of the monopole and quadrupole moments of the 2D correlation function. We show the covariance matrices in Fig.~\ref{covs} for the same sequence of models at redshift $z = 0.3915$, but now additionally include the noise variance in our calculation, corresponding to an approximate total integration time of 2150 hours ($\bar n \approx 10^{-3}$ Mpc$^{-3}$). We use a range of separations from 40--190 Mpc for the monopole and 80--190 Mpc for the quadrupole, with separation bins of $\Delta r = 2$ Mpc. 

In the case of the base cosmology (first panel), only covariance blocks $\ell=\ell^\prime$ are non-zero in accordance with there being no anisotropic effects present. The covariance is larger at smaller separations (the variance of the 21cm field is larger on smaller scales), and there is a moderately broad band around the diagonal for both the monopole and quadrupole, indicating the correlation between neighbouring separation bins.

When RSDs are included (second panel), a number of significant changes occur. First, a large anti-correlation arises in a substantial fraction of the \{0,2\} block. The magnitude of the covariance is increased in general, particularly in the quadrupole-quadrupole (\{2,2\}) block. These changes can be understood analytically; at a given redshift, and neglecting the Fingers of God effect, the multipoles of $P_{\text{RSD}}$ are multiplicative constants, determined by the values of the bias and growth factor. In the covariance expression (Eq.~\ref{eq:multipolecov}), we take sums over such factors. The coefficients for the RSD that appear in the monopole and particularly the quadrupole are quite large ($c_{2,\text{RSD}} \approx 6$; c.f. \cite{Tansella2018}), hence the substantial enhancement of the corresponding covariance matrix elements.

\begin{figure}
    {\centering
	\includegraphics[width=0.99\columnwidth]{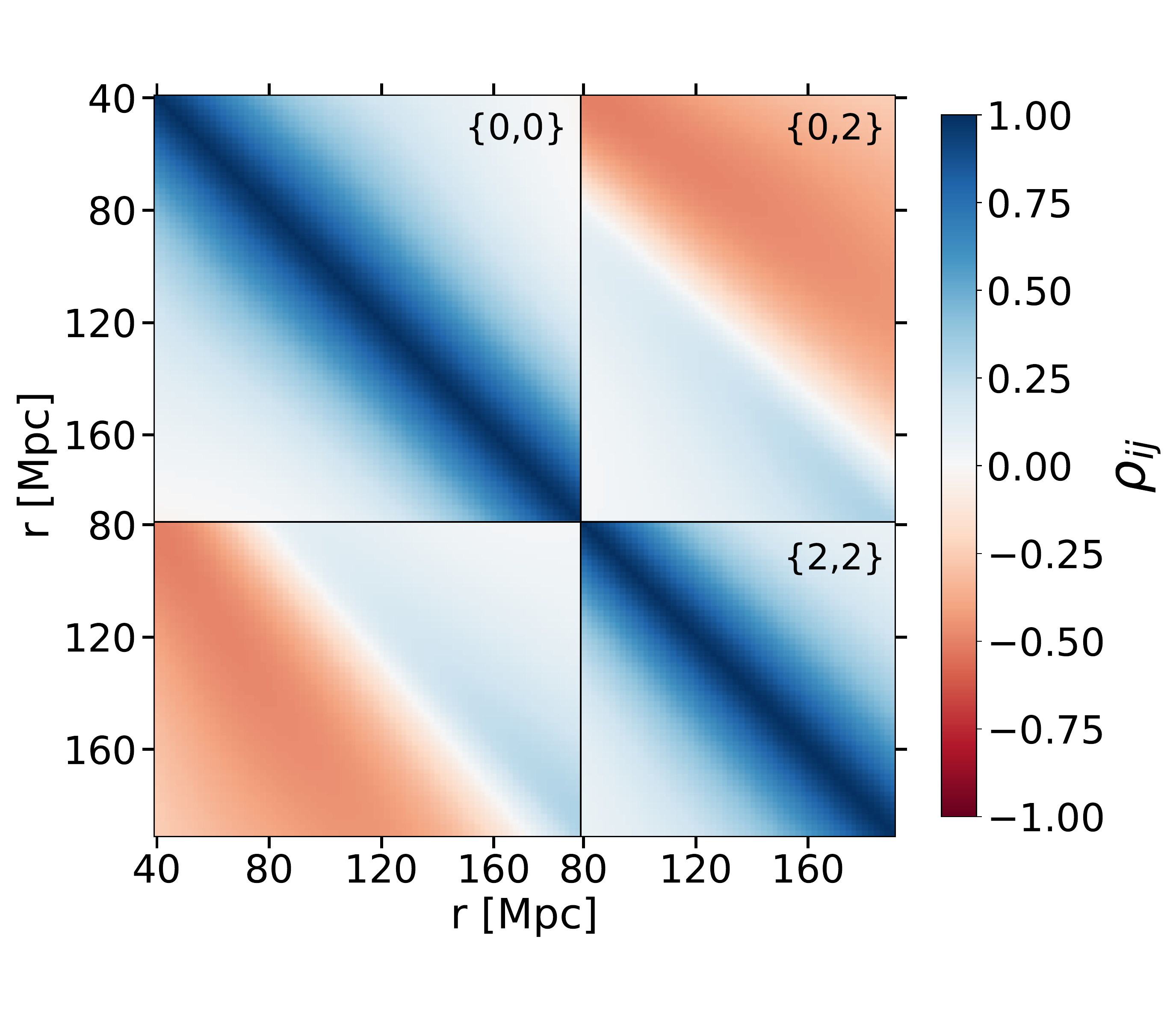}
	\includegraphics[width=0.99\columnwidth]{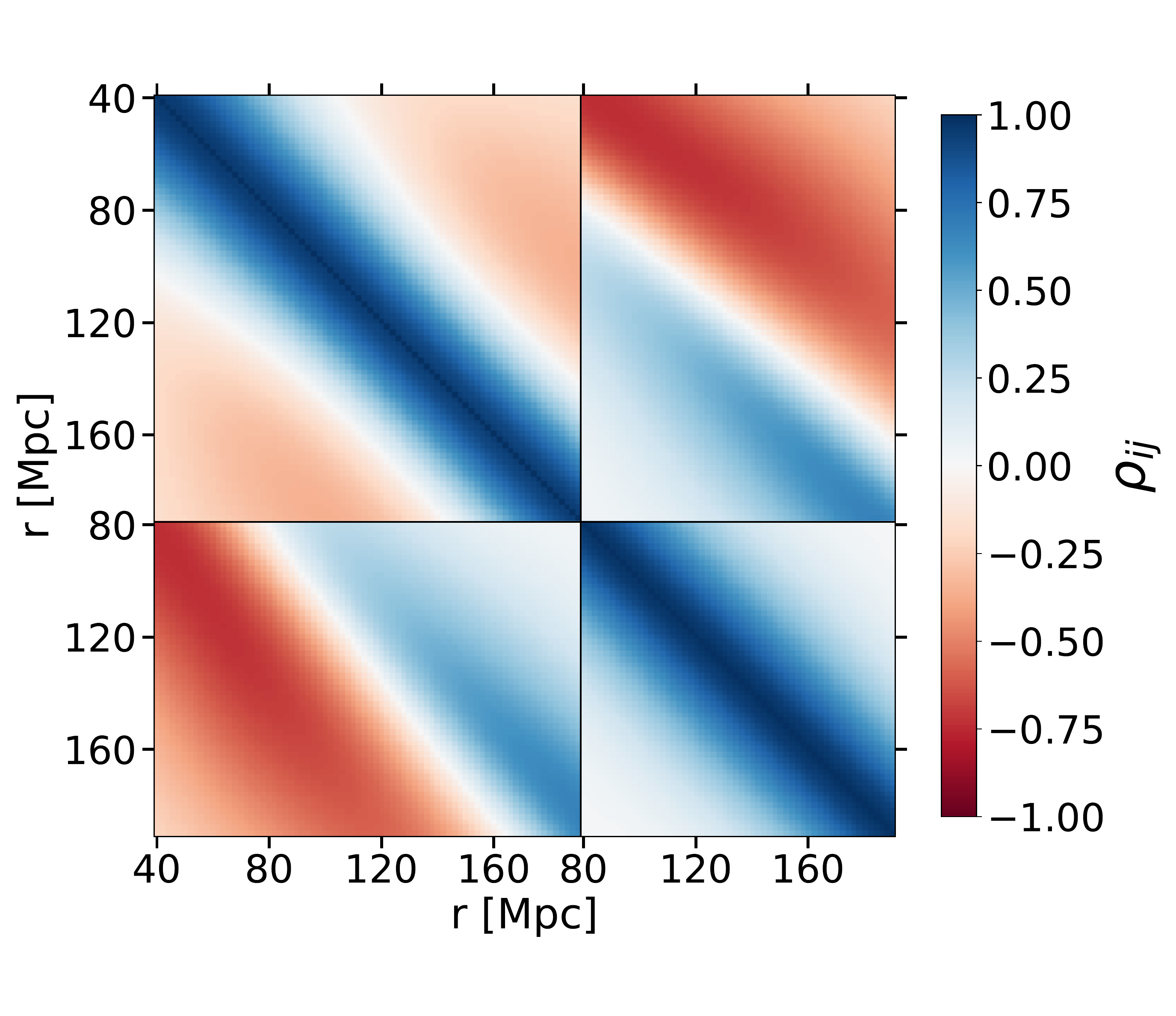}}
	\caption{Correlation matrices, $\rho_{ij} = C_{ij}/\sqrt{C_{ii}C_{jj}}$, for two different scenarios. {\it (Upper panel):} Base cosmology + RSDs, with no beam smoothing or foreground cut. {\it (Lower panel):} Same as above, but now with the fiducial beam and foreground cut included. Note the enhanced monopole-quadrupole (anti-)correlations in the latter case.}
	\label{corrcovs}
\end{figure}

At this point, we note that we have validated our covariance matrix calculations against the {\tt COFFE} code \citep{Tansella2018}. We performed our comparisons at $z = 1$ with matching input power spectra and RSD coefficients, recovering the {\tt COFFE} result to within 0.1\% in the vicinity of the diagonal, with a sub-1\% residual elsewhere (outside of zero-crossings). We expect that this residual is due to the different numerical integration scheme implemented in the {\tt COFFE} code, and do not expect it to significantly affect our results.

The third panel of Fig.~\ref{covs} adds a beam function into the covariance calculation. 
Its main effect is to attenuate the covariance in the $\ell=\ell^\prime$ blocks, i.e. it reduces the amplitude of the covariance matrix elements. This is consistent with the fact that the beam acts to smooth the 21cm fluctuation field, reducing its overall variance and preferentially destroying small scale information (i.e. at separations below the beam scale, the field becomes strongly correlated, but its variance is suppressed).

In the final case of the addition of the foreground cut (fourth panel), additional attenuation is observed, particularly for the monopole-monopole (\{0,0\}) block. An anti-correlation is also introduced into the off-diagonal region of this block, which can be seen more clearly in Fig.~\ref{corrcovs}. This is most likely related to how the foreground cut changes the amplitude and shape of the smooth continuum part of the correlation function, which is a non-local effect in separation.

In Fig.~\ref{corrcovs} we additionally plot correlation matrices for two cases: one with no systematics present (only the isotropic + RSD components), and another with beam smoothing and a foreground cut also included. The strength of correlations and anti-correlations is much larger in the \{0,2\} block when including the beam and foreground cut, and (as mentioned above) an anti-correlated region is introduced into the off-diagonal part of the \{0,0\} block. 

\begin{figure}
     \includegraphics[scale=0.20]{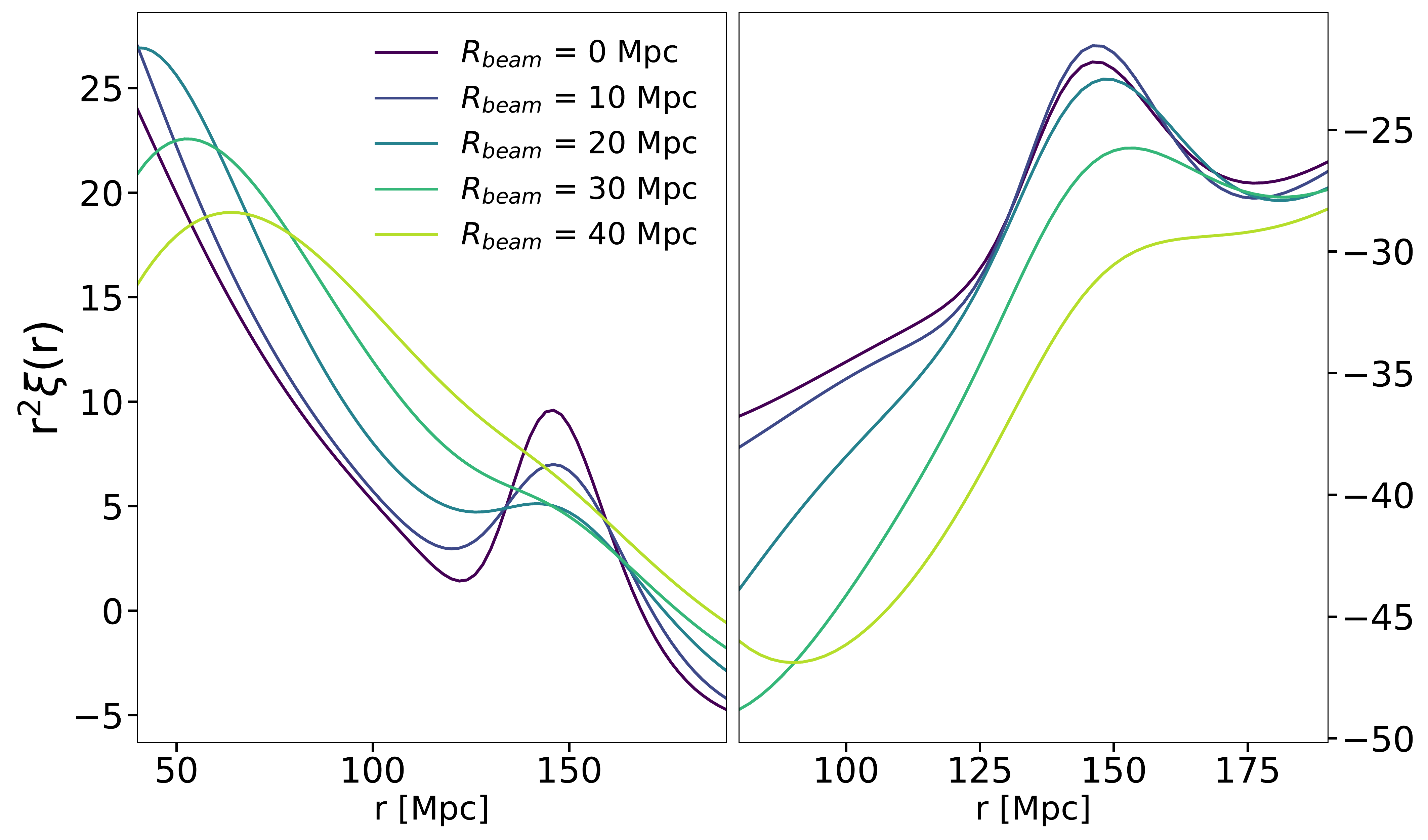}\vfill
    \includegraphics[scale=0.20]{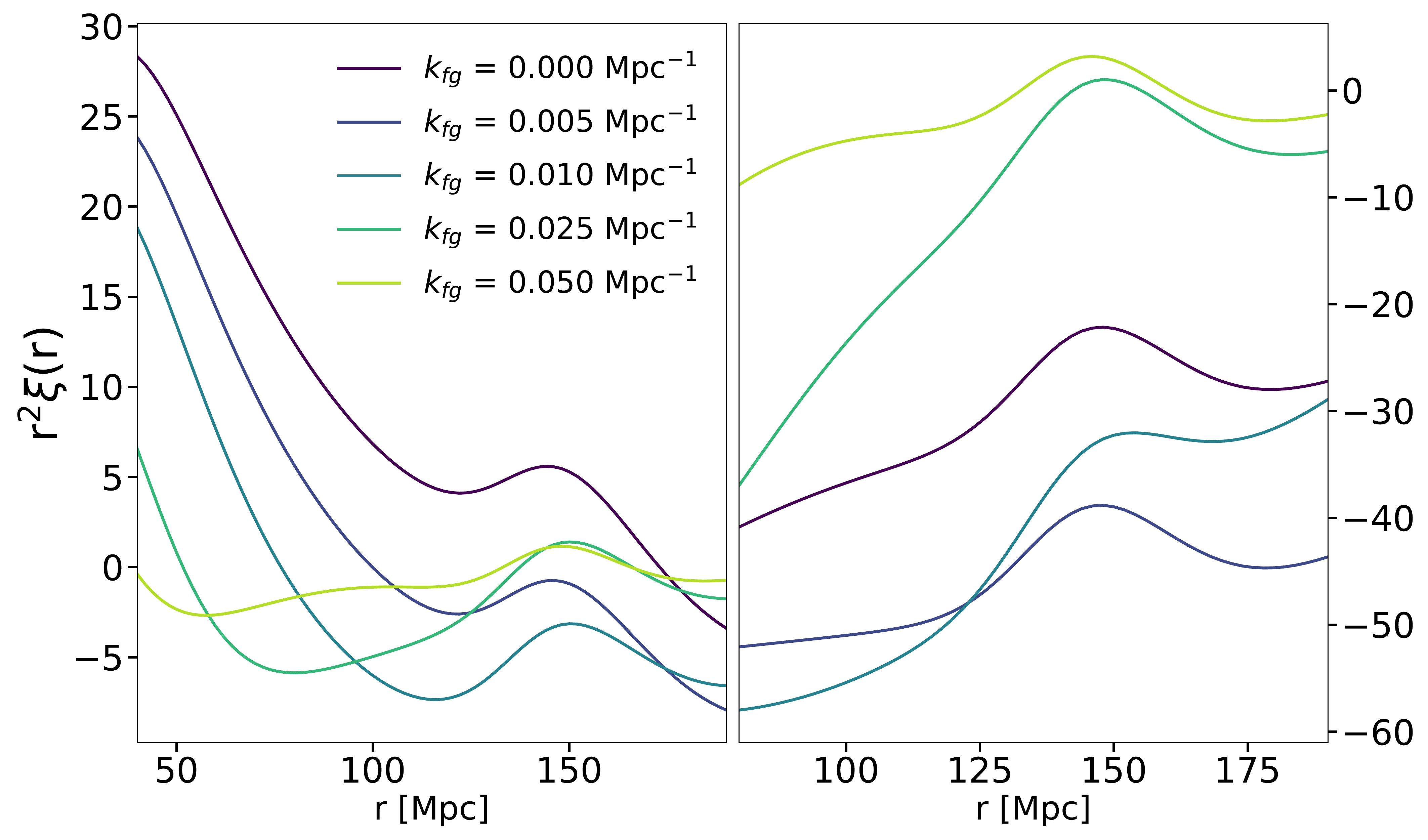}
    \caption{The monopole (left) and quadrupole (right) with increasing Gaussian beam width (top) and increasing foreground cut (bottom). The BAO feature in the quadrupole exhibits less smoothing as $R_{\text{beam}}$ increases. Foreground $k_{\parallel}$ filtering dramatically changes the shape of the continuum, and also slightly alters the position of the BAO peak.}
    \label{models}
\end{figure}

\subsection{Effect of beam smoothing and foreground cuts}

In Fig.~\ref{models} we show how different levels of beam smoothing and foreground cuts affect the monopole and quadrupole of the 2D correlation function. The range of separations chosen for fitting the quadrupole ($80-190$ Mpc) does not contain the lowest separations from the monopole region ($40-190$ Mpc) due to the added complexity of fitting it in this range. The upper panels show the effect of applying Gaussian beams of differing comoving widths $R_{\rm beam}$ (see Eq.~\ref{eq:gaussian_beam} for a definition). RSDs are included in these calculations, but a foreground cut is not. The effect of the beam is similar to the one studied in \cite{Villaescusa-Navarro2017} in the monopole case, where the angle-averaged BAO feature is smoothed out as the beam width increases, becoming essentially indistinguishable from the underlying continuum beyond $R_{\rm beam}\approx 40$ Mpc. Referring back to Fig.~\ref{fig:beamsize}, this corresponds to $z \approx 0.8$ for MeerKAT, implying that the BAO scale cannot be recovered from the monopole of the correlation function beyond this redshift.

In the case of the quadrupole, increasing the beam width also increasingly smooths-out the BAO feature, but to a lesser extent than in the monopole, and in fact the BAO feature remains well-defined at $R_{\rm beam} = 30$ Mpc. This is a result of the down-weighting of the beam-suppressed transverse directions in the 2D correlation function by the quadrupole. Additional BAO information can also be extracted from higher multipoles, although these are increasingly noisy compared to the monopole and quadrupole. 

In the lower panel of Fig.~\ref{models}, the effect of an increasingly severe foreground cut, $k_{\rm fg}$, is shown. RSDs are again included in each case, as is a beam smoothing with $R_{\rm beam} = 16.9$ Mpc. As $k_{\rm fg}$ increases, the monopole of the correlation function is pulled down to smaller and smaller amplitudes, but without much change in the sharpness of the BAO feature. This continues until around $k_{\rm fg} \approx 0.02$ Mpc$^{-1}$, when the amplitude begins to increase again, the shape of the correlation function around the BAO scale begins to change, and the BAO peak begins to be suppressed. The latter behaviour can be understood as being due to the foreground cut starting to eat into radial modes at which the BAO wiggles are present in the power spectrum, $k \gtrsim 0.2$ Mpc$^{-1}$, therefore destroying some of the available BAO information. Before this point, the foreground cut primarily removes low-$k$ modes that mostly only affect the continuum of the correlation function. A similar pattern is also seen for the quadrupole, with large changes in amplitude but smaller modifications to the shape of the correlation function as $k_{\rm fg}$ is increased.

\subsection{Model fitting under different conditions}

In this section, we study the effectiveness of the model-fitting procedure described in Sect.~\ref{sec:fittingmethod} as the various anisotropic effects are included in the model (Sect.~\ref{sec:anisotropic}), and as various analysis assumptions are changed: the thermal noise level (Sect.~\ref{sec:thermalnoise}); the assumed beam model (Sect.~\ref{sec:beamcomp}); the extent of the foreground cut (Sect.~\ref{sec:fgcuts}); and whether the beam assumed in the covariance calculation matches the true one (Sect.~\ref{sec:nonoptcov}).

\subsubsection{Combinations of anisotropic effects}
\label{sec:anisotropic}

In this section, we show the results of least-squares fits of the correlation function model defined in Sect.~\ref{sec:fittingmethod} in terms of the recovered values of the radial and transverse $\alpha$ parameters, for simulated data containing different combinations of anisotropic effects.

\begin{figure}
	\includegraphics[scale=0.23]{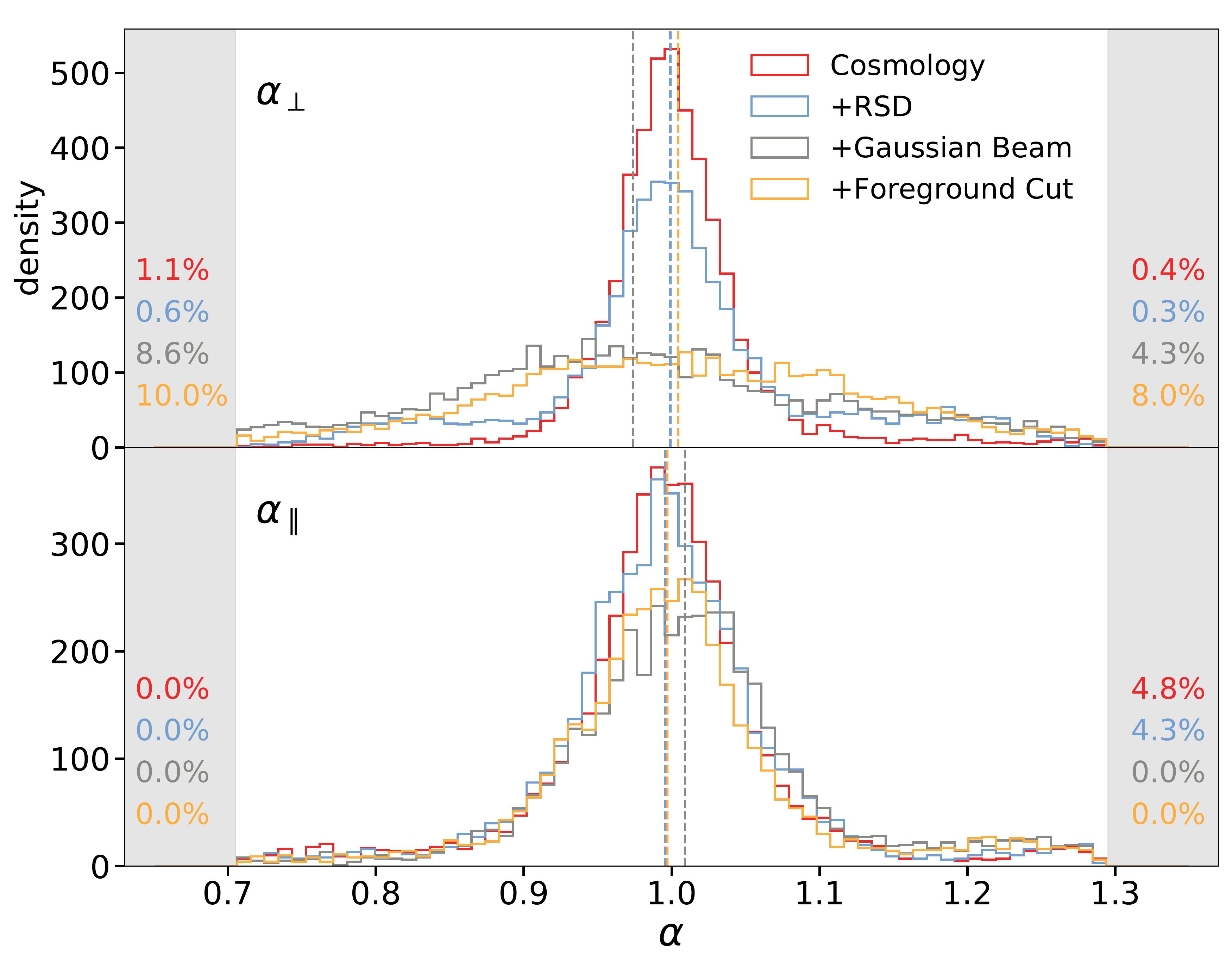}
	\caption{Recovered $\alpha$ parameter values from fits to 5000 realisations of a series of models, each adding an additional anisotropic effect to the last, starting with the fiducial configuration at $z = 0.3915$. Percentages of catastrophic fits are listed in the grey edge bands at either side. The vertical dashed lines show the median of the non-catastrophic fits. The $\alpha_{\perp}$ distribution is significantly flattened after addition of the beam.}
	\label{contribs-dist}
\end{figure}
\begin{table}
\centering
\begin{tabular}{ |c|c|c|c|c|c|  }
 \hline
 Run & & $\Delta$Med. &  $\Delta$Mean & $\sigma$ & MAD\\
 \hline
 Cosmology & $ \alpha_{\perp}    $ & -0.001 & 0.004 & 0.058 & 0.024 \\
 & $ \alpha_{\parallel}$ & -0.004 & -0.004 & 0.079 & 0.034 \\
 \hline
 RSD added & $ \alpha_{\perp}    $ & -0.001 & 0.004 & 0.095 & 0.038 \\
 & $ \alpha_{\parallel}$ & -0.005 & -0.002 & 0.080 & 0.038 \\
 \hline
 Beam added & $ \alpha_{\perp}    $ & -0.026 & -0.019 & 0.127 & 0.083 \\
 & $ \alpha_{\parallel}$ & 0.009 & 0.013 & 0.089 & 0.044 \\
 \hline
 FG cut added & $ \alpha_{\perp}    $ & 0.005 & 0.006 & 0.121 & 0.086 \\
 & $ \alpha_{\parallel}$ & -0.003 & 0.002 & 0.088 & 0.039 \\
 \hline
\end{tabular}
\caption{Statistics from runs including sequential additions of anisotropic effects in Fig.~\ref{contribs-dist}. Catastrophic fits have been removed.
} \label{tab-contribs}
\end{table}

The configuration including all of the effects -- RSDs, beam smoothing, and the foreground cut -- is adopted as our fiducial model throughout the rest of the paper, with relevant parameters set to the following values: \{$R_{\text{beam}} = 16.9\,\text{Mpc}$, $k_{\text{fg}} = 0.01$ Mpc$^{-1}$, $\bar n_{\rm IM} = 3.74\times 10^{-3}$ Mpc$^{-3}$, $f_{\text{sky}} = 0.1$\}, all in the redshift band centred at $z=0.3915$. A Gaussian model is used for the beam in both the simulated data and the fitting function, and the beam width, $R_{\rm beam}$, is treated as a free parameter. The choice was made to use the Gaussian beam model rather than the model from the {\tt katbeam} package for the fiducial case because the two give very similar covariance matrices; the correlation structure is unchanged, and individual elements differ by less than 1\% in the vicinity of the diagonal. The true beam is also known imperfectly, to within a few percent, and so the calculated difference in our model covariances is smaller than the accuracy to which the beam is known. Furthermore, the Gaussian beam model has the advantage of allowing for quicker evaluation of model fits, and using it offers an opportunity to study interactions between the beam width parameter $R_{\text{beam}}$ and other fitting parameters. The full continuum model from Eq.~\ref{eq:continuum} is included, with all parameters allowed to vary. We do not allow the parameters of the RSD model to vary however, and we fix the amplitude of the BAO feature to $A=1$. 
The fits were performed on 5000 Gaussian random realisations of the monopole and quadrupole correlation functions at separations of 40--190 Mpc and 80--190 Mpc respectively, with separation bins of width $\Delta r = 2$ Mpc. The realisations are generate from the corresponding `true' correlation function model and covariance matrix in each case.

The distributions of the recovered $\alpha$ values are plotted in Fig.~\ref{contribs-dist}, while Table~\ref{tab-contribs} shows summary statistics for the distributions. The summary statistics include the difference between the expected median and mean (unity in each case), denoted as $\Delta$med. and $\Delta$Mean respectively; the standard deviation of the distribution, $\sigma$; and the median absolute deviation (MAD) of the distribution, which is more robust to outliers than $\sigma$. All statistics are calculated after removing catastrophic errors, which are defined as any recovered $\alpha$ values that hit the edge of the allowed prior range. The percentage of fits removed after hitting each prior edge is shown on each side of the figure.

In the fits we use a
5$\%$ prior range about the true value of $R_{\text{beam}}$ and a prior range on the $\alpha$ parameters of \{0.7,1.3\}. In each of the 4 runs shown, an additional anisotropic effect is included on top of the ones already included in the previous case. For clarity, we reiterate that each set of simulations was generated using a covariance matrix including the set of anisotropic effects pertinent to that case.

\begin{figure}
	\includegraphics[scale=0.23]{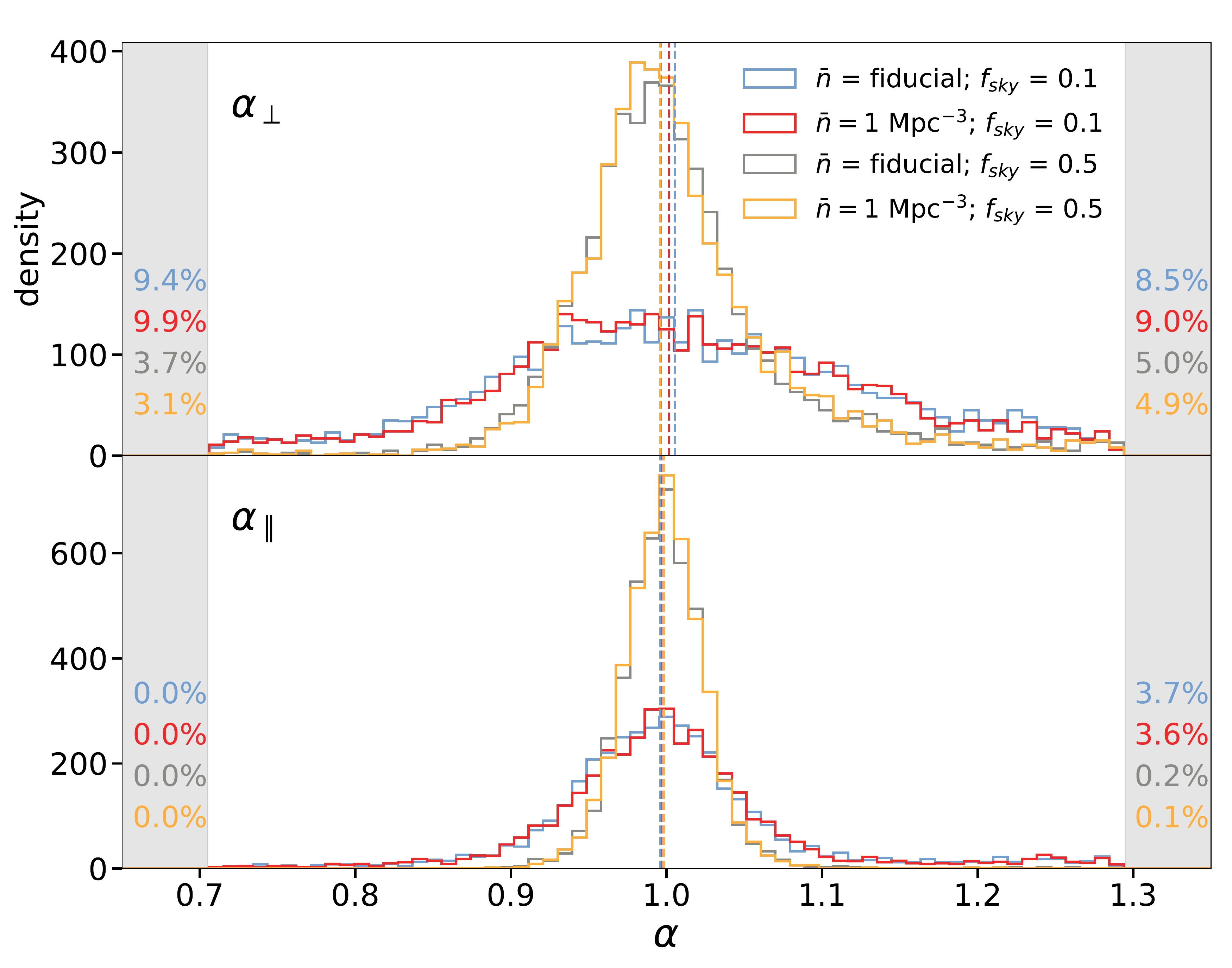}
	\caption{Recovered $\alpha$ parameters from fits to 5000 realisations for different noise level and survey areas. The fiducial noise value $\bar n = 3.74 \times 10^{-3}$ Mpc$^{-3}$ is close to the sample-variance limit, which is a consequence of how the survey area has been optimised for the MeerKLASS survey.}
	\label{noise-dist}
\end{figure}
\begin{table}
\centering
\begin{tabular}{ |c|c|c|c|c|c|  }
 \hline
 Run & & $\Delta$ Med. &  $\Delta$Mean & $\sigma$ & MAD\\
 \hline
  $\bar n$ = fiducial; $f_{\text{sky}} = 0.1$ & $ \alpha_{\perp}$ & 0.005 & 0.010 & 0.118 & 0.078 \\
 & $ \alpha_{\parallel}$ & -0.004 & 0.002 & 0.081 & 0.036 \\
 \hline
  $\bar n = 1$ Mpc$^{-3}$; $ f_{\text{sky}} = 0.1$ & $ \alpha_{\perp}$ & 0.002 & 0.008 & 0.115 & 0.074 \\
  & $ \alpha_{\parallel}$ & -0.003 & 0.002 & 0.080 & 0.035 \\ 
  \hline
  $\bar n = $ fiducial; $f_{\text{sky}} = 0.5$& $ \alpha_{\perp}$ & -0.004 & 0.005 & 0.070 & 0.033 \\ 
 & $ \alpha_{\parallel}$ & -0.002 & -0.001 & 0.030 & 0.017 \\\hline
  $\bar n = 1$ Mpc$^{-3}$; $f_{\text{sky}} = 0.5$& $ \alpha_{\perp}$ & -0.004 & 0.006 & 0.069 & 0.032 \\ 
 & $ \alpha_{\parallel}$ & -0.002 & -0.001 & 0.029 & 0.016 \\ 
 \hline
\end{tabular}
\caption{Statistics from runs at different noise levels in Fig.~\ref{noise-dist}. Catastrophic fits removed.}
\label{tab-noise}
\end{table}


From Fig.~\ref{contribs-dist}, we see that the width of the $\alpha_{\perp}$ distribution increases significantly upon the introduction of beam smoothing, but does not cause the same change in the line-of-sight parameter, $\alpha_{\parallel}$. Despite fitting for the beam width, its introduction results in a bias in the median value of $\alpha_{\perp}$ of around 2$\%$, although this bias disappears on the introduction of the foreground cut. We study the effect of different foreground cut values further in Sect.~\ref{sec:fgcuts}.

These results are a consequence of the strong smoothing of the BAO feature in the transverse direction that was shown in Fig.~\ref{2dcorrs}. For this particular MeerKAT-like survey configuration, it is clear that $\alpha_{\perp}$ will be difficult to recover due to the beam, while the recovery of $\alpha_{\parallel}$ would face only slightly more difficulty than in the case of a galaxy survey configuration over the same survey volume. This lends further support to the proposal for making use of only the line-of-sight power spectrum in \citet{Villaescusa-Navarro2017}.

Another feature of interest is the slight asymmetry of each of the $\alpha_{\perp}$ distributions, with a larger tail into the $\alpha > 1$ region, and a median value just greater than unity even in the simplest case of a base cosmological power spectrum only. Due to the presence of these tails, we include the median absolute deviation of each distribution, ${\rm MAD}(\alpha) \equiv {\rm Med}(|\alpha - {\rm Med}(\alpha)|)$, in our results tables as a separate comparison of the distribution width that is more robust to non-Gaussian tails. The likely reason for these tails is overfitting and partial degeneracies with the continuum component of the fitting model. Note that we did study alternative forms for the continuum models, but found the one in Eq.~\ref{eq:continuum} to perform best in our tests.

We note that the width of the recovered $\alpha$ distributions is quite large even in the absence of the beam smoothing and foreground cut. As we will show in the next section, this is largely due to the fiducial MeerKAT survey configuration that we are considering (with $f_{\rm sky} = 0.1$) saturating the sample variance bounds.

\begin{figure}
	\includegraphics[scale=0.23]{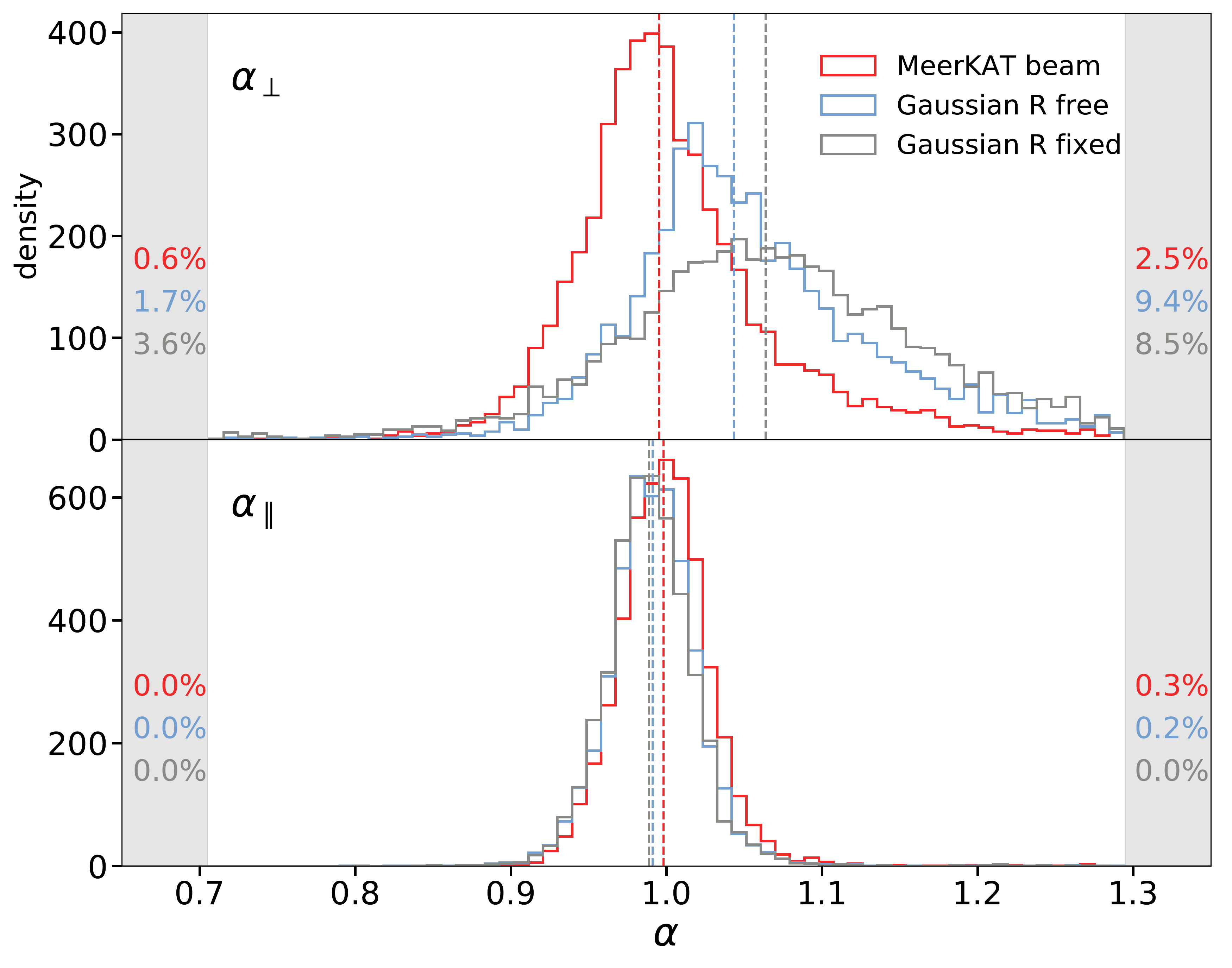}\hfill
	\caption{Recovered $\alpha$ parameters from fits to realisations generated using the MeerKAT beam model at $z=0.3915$ under ideal noise conditions ($\bar n = 5\times10^{-3}$ Mpc$^{-3}$, $f_{\text{sky}} = 0.5$). Fitting a Gaussian beam to data generated using the {\tt katbeam} model introduces significant bias to $\alpha_{\perp}$. }
	\label{MKvG-dist}
\end{figure}

\subsubsection{Thermal noise level}
\label{sec:thermalnoise}

In Fig.~\ref{noise-dist} and Table~\ref{tab-noise} we show the distributions of the recovered $\alpha$ parameters under the fiducial conditions set out in Sect.~\ref{sec:anisotropic} with all anisotropic effects included, but now with changes to the noise and survey area parameters, $\bar{n}_{\rm IM}$ and $f_{\text{sky}}$, in the covariance matrix. In the case that the effective $\bar{n}_{\rm IM}$ is made approximately 300 times larger than the fiducial value of $3.74\times10^{-3}$ Mpc$^{-3}$ but $f_{\rm sky}$ remains fixed, the distributions for both $\alpha_{\perp}$ and $\alpha_{\parallel}$ show very little difference. This suggests that our fiducial value of $\bar{n}_{\rm IM}$ is close to the sample variance limit, which is to be expected given that the MeerKLASS survey area has been optimised for a BAO detection.

Increasing the volume of the survey via $f_{\text{sky}}$ makes a much more substantial difference to the recovery of the BAO scale, regardless of whether $n_{\rm IM}$ remains fixed or is increased. In particular, setting $f_{\rm sky} = 0.5$ substantially reduces the width of the $\alpha_\parallel$ distribution, from around 8\% to 3\%, as well as decreasing the width of the $\alpha_\perp$ distribution from around 12\% to 7\%, despite the presence of the beam smoothing and foreground cut. This case also reveals again the non-Gaussian, positive-tailed shape of the $\alpha_{\perp}$ distribution compared to $\alpha_{\parallel}$. While a survey area of $f_{\rm sky} = 0.5$ is likely out of reach of MeerKAT, The SKAO Mid telescope is expected reach a similar $\bar{n}_{\rm IM}$ to the MeerKAT configuration that we study over an area approaching this value. 

\subsubsection{MeerKAT beam versus Gaussian approximation}
\label{sec:beamcomp}

In Fig.~\ref{MKvG-dist} and Table~\ref{tab-MKvG} we show the results of fitting the correlation function multipoles under three different beam assumptions to 5000 random realisations, now generated using the {\tt katbeam} model, which we consider to be the `true' MeerKAT beam. Additionally in these runs, the covariance used to generate the realisations assumes a larger survey area ($f_{\text{sky}} = 0.5$), to ensure that the beam model is the dominant factor in the performance of the fits.

\begin{figure}
	\includegraphics[scale=0.23]{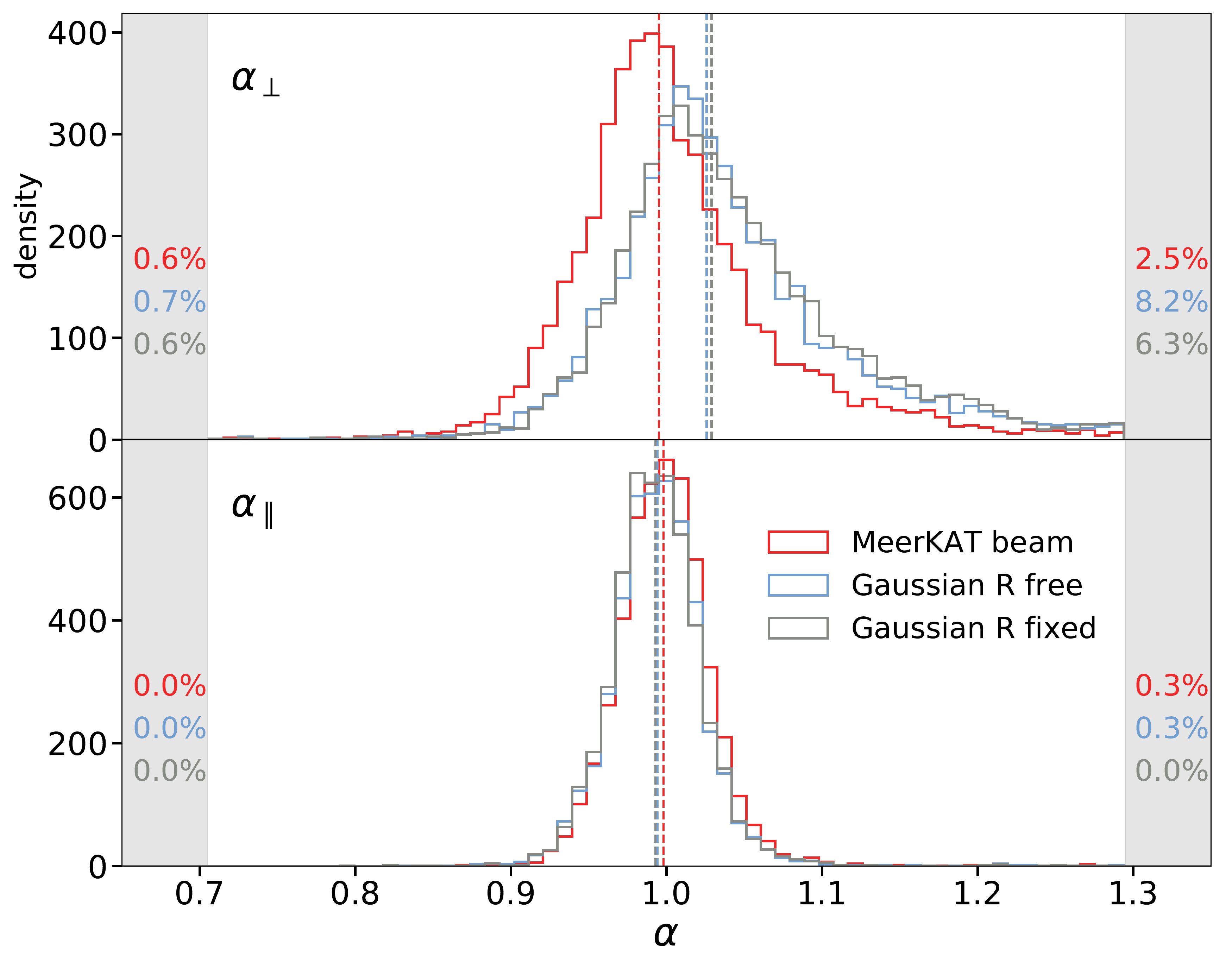}\hfill
	\caption{Recovered $\alpha$ parameters from fits to MeerKAT beam realisations with a Gaussian $R_{\text{beam}} = 15.72$ Mpc, matched to the {\tt katbeam} Hankel Transform FWHM rather than the real-space counterpart. }
	\label{MKvG-dist2}
\end{figure}

\begin{table}
\centering
\begin{tabular}{ |c|c|c|c|c|c|  }
 \hline
 Run & & $\Delta$Med. &  $\Delta$Mean & $\sigma$ & MAD\\
 \hline
 {\tt katbeam} & $ \alpha_{\perp}    $ & -0.005 & 0.005 & 0.067 & 0.033 \\
 & $ \alpha_{\parallel}$ & -0.002 & -0.001 & 0.032 & 0.018 \\
  \hline
 $R_{\text{b}}$ free  & $ \alpha_{\perp}    $ & 0.043 & 0.054 & 0.078 & 0.044 \\
 & $ \alpha_{\parallel}$ & -0.009 & -0.008 & 0.031 & 0.017 \\ \hline
 $R_{\text{b}}$ fixed & $ \alpha_{\perp}    $ & 0.064 & 0.065 & 0.094 & 0.059 \\
 & $ \alpha_{\parallel}$ & -0.011 & -0.010 & 0.031 & 0.017 \\
 
 \hline
 \hline

  $R_{\text{b}}$ free (HT) & $ \alpha_{\perp}    $ & 0.026 & 0.038 & 0.073 & 0.038 \\
 & $ \alpha_{\parallel}$ & -0.006 & -0.005 & 0.032 & 0.017 \\ \hline
  $R_{\text{b}}$ fixed (HT)& $ \alpha_{\perp}    $ & 0.029 & 0.042 & 0.074 & 0.040 \\
 & $ \alpha_{\parallel}$ & -0.007 & -0.006 & 0.032 & 0.017 \\
 \hline
\end{tabular}
\caption{Statistics from testing recovery after changes to the beam model in Figures \ref{MKvG-dist} and \ref{MKvG-dist2}; Gaussian models parametrised by $R_{\text{beam}}$ matched to real-space FWHM, and the Hankel transform (HT) FWHM. Catastrophic fits removed.}
\label{tab-MKvG}
\end{table}

\begin{figure*}
	\includegraphics[width=\columnwidth]{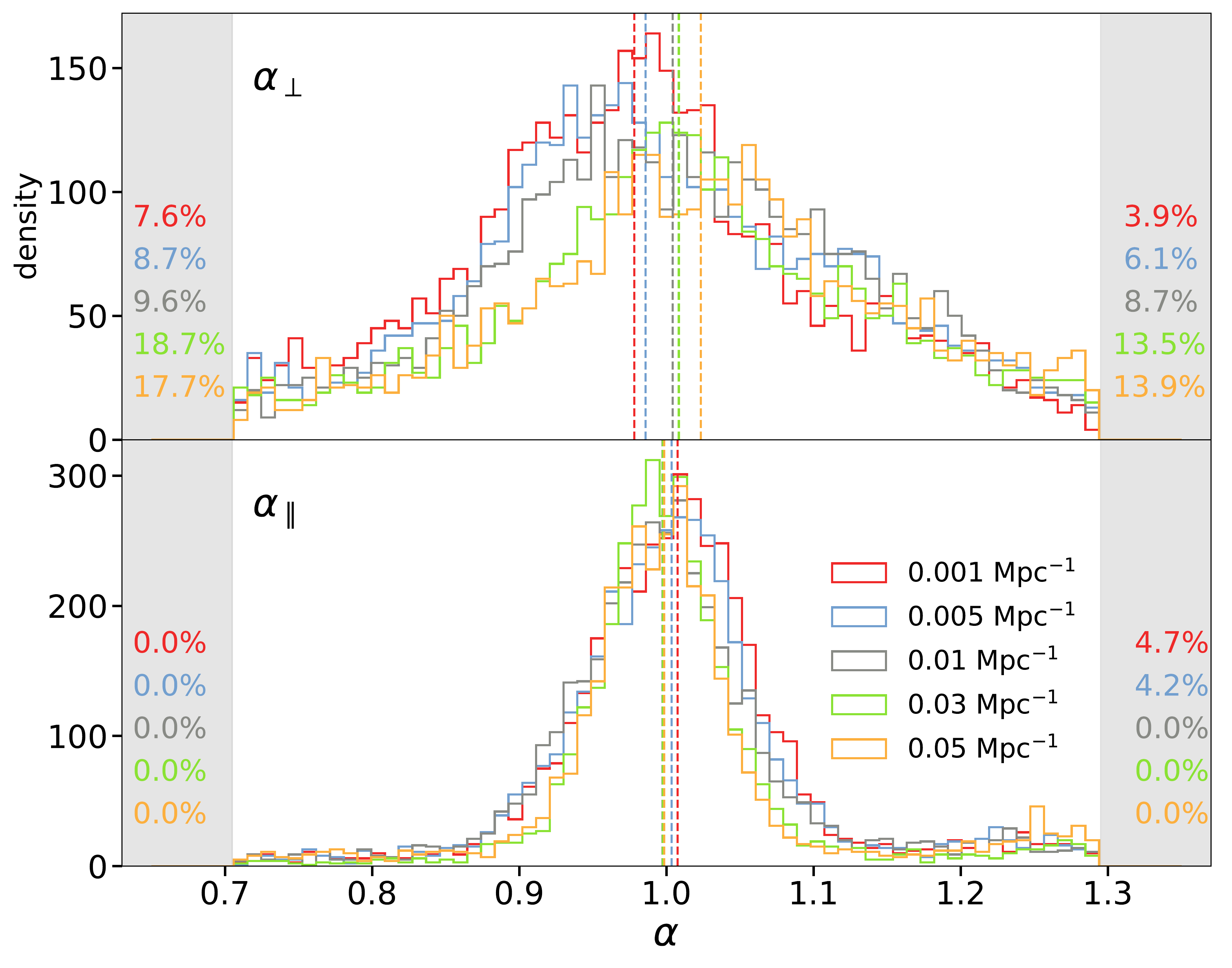}
	\includegraphics[width=\columnwidth]{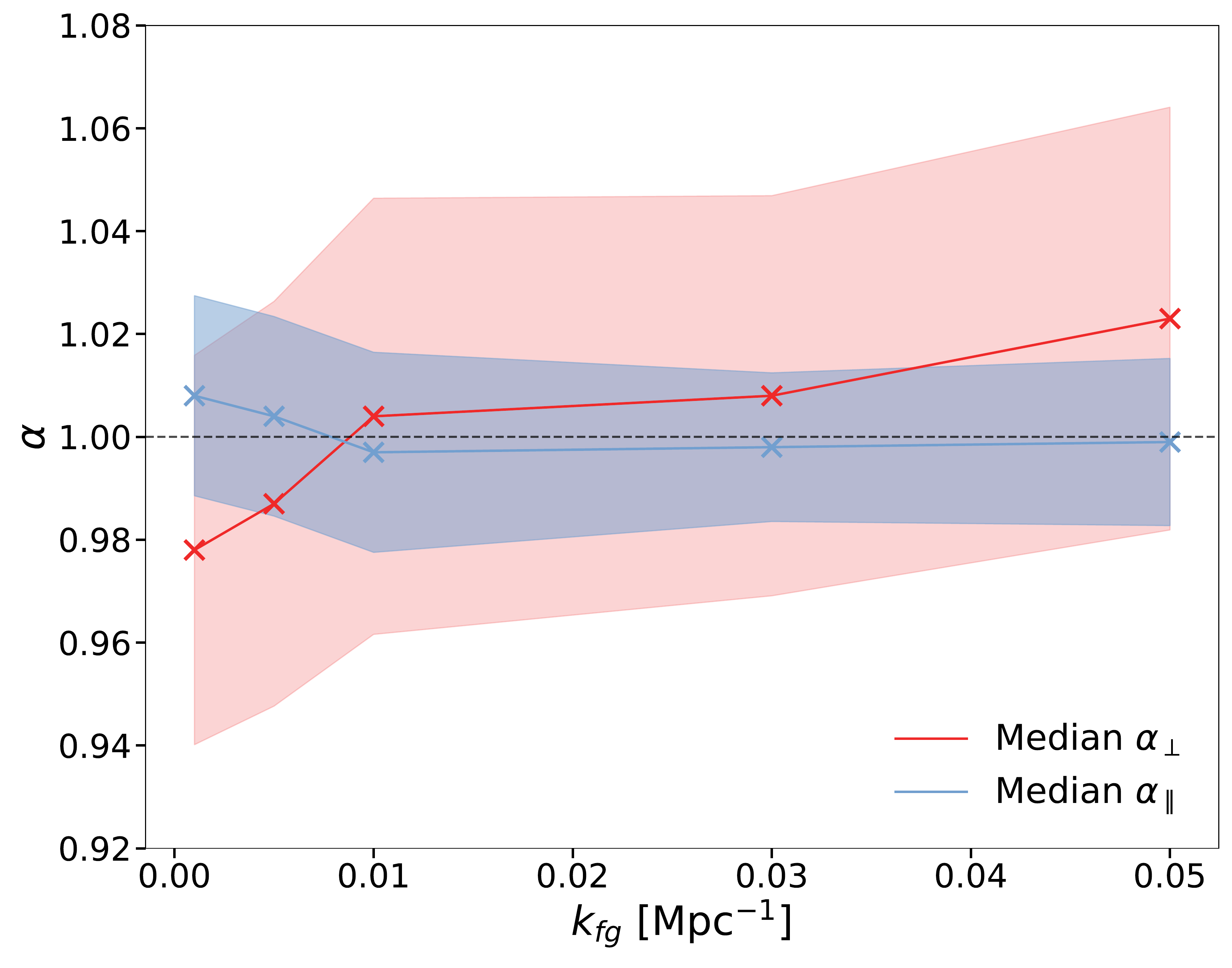}
	\caption{{\it Left panel:} Recovered $\alpha$ parameters from fits to 5000 realisations at $k_{\text{fg}}$ values shown. {\it Right panel:} Change in the median of both $\alpha$ parameters with foreground cut, with overplotted color bands showing the median absolute deviation of each point. The $\alpha$ parameters show bias on opposite sides of the fiducial line, which changes with the application of greater foreground cuts.}
	\label{kfg-dist}
\end{figure*}

We consider three scenarios for the fitting models: one with a fixed `true' MeerKAT beam from {\tt katbeam}; one with a Gaussian beam with $R_{\rm beam}$ allowed to vary (as in previous sections, with a 5\% prior); and one with a Gaussian beam fixed so that its effective FWHM matches the {\tt katbeam} FWHM. Even when fitting the MeerKAT beam to itself, the distribution of the $\alpha_{\perp}$ parameter is not symmetric, with a larger tail into positive values, suggesting some inherent difficulty in fitting $\alpha_{\perp}$, perhaps due to over-fitting or partial degeneracies with the continuum model.
The runs with a Gaussian beam, both fixed and allowed to vary, show large biases to the median of $\alpha_{\perp}$ of 5.4\% and 6.4\% respectively. The $\alpha_{\parallel}$ parameter is biased negative at the 1\% level in both of these cases, likely compensating slightly for the increase in the $\alpha_{\perp}$ parameter, although the effect is small. From this we can conclude that the $\alpha_{\parallel}$ distribution is quite stable to the assumed beam model -- an incorrect beam will mostly only impact the recovery of the transverse BAO parameter.

To understand the reasons for the large bias appearing when using the Gaussian model, we also examined the beam functions themselves and their Hankel transforms. Though the fixed Gaussian beam model matches the FWHM of the {\tt katbeam} output in real space, when both functions are Hankel transformed, the functions are not well-matched in width, due to the extra structure in the {\tt katbeam} model at wider angles (e.g. sidelobes). This motivated a further comparison with the Gaussian beam and {\tt katbeam} models matched at their FWHM in Fourier space instead, which corresponded to a Gaussian beam width of $R_{\text{beam}} = 15.72$ Mpc.

In Fig.~\ref{MKvG-dist2} and the lower section of Table~\ref{tab-MKvG}, we show the results of runs using this assumption instead, again for a free beam width with a 5\% prior, and a fixed beam width that is now set to the Hankel transform value. We find that the fixed $R_{\rm beam}$ model does indeed offer a slight improvement over the case where the FWHM was matched for the real-space beams. In the fixed beam case, the median bias on $\alpha_{\perp}$ decreases from 6.4\% to 2.9\%, while for free $R_{\text{beam}}$ it decreases from 4.3\% to 2.6\%. The former result is due to the better match of the beam smoothing functions in Fourier space, while the latter is most likely due to the shift in the prior range of $R_{\rm beam}$. Nevertheless, a bias remains in all of the Gaussian cases that is not seen when the true ({\tt katbeam}) model is used, suggesting that the detailed shape of the beam is a material factor in the analysis, even if the median bias is smaller than one standard deviation.

Additionally, we note that the width of the recovered distributions for $\alpha_\perp$ did not change much between the fixed-width and variable-width Gaussian cases once the Hankel transform FWHM was adopted, with both having essentially identical values for both $\sigma$ and the MAD. This suggests that allowing the beam width to be a free parameter does not significantly degrade the measurement precision on either $\alpha_\perp$ or $\alpha_\parallel$, and so there should be no reason not to marginalise over this parameter in analyses.


\begin{table}
\centering
\begin{tabular}{ |c|c|c|c|c|c|  }
 \hline
 $k_{\text{fg}}$ [Mpc$^{-1}$] & & $\Delta$Med. &  $\Delta$Mean & $\sigma$ & MAD\\
 \hline
  
 0.001 & $ \alpha_{\perp}    $ & -0.022 & -0.019 & 0.121 & 0.076 \\
 & $ \alpha_{\parallel}$ & 0.008 & 0.010 & 0.081 & 0.039 \\\hline
 0.005 & $ \alpha_{\perp}    $ & -0.014 & -0.004 & 0.124 & 0.082 \\
 & $ \alpha_{\parallel}$ & 0.004 & 0.007 & 0.085 & 0.039 \\\hline
 0.01 & $ \alpha_{\perp}    $ & 0.004 & 0.007 & 0.122 & 0.085 \\
 & $ \alpha_{\parallel}$ & -0.003 & 0.002 & 0.085 & 0.039 \\\hline
 0.03 & $ \alpha_{\perp}    $ & 0.008 & 0.011 & 0.125 & 0.078 \\
 & $\alpha_{\parallel}$ & -0.002 & 0.005 & 0.073 & 0.029 \\\hline
 0.05 & $ \alpha_{\perp}    $ & 0.023 & 0.021 & 0.127 & 0.082 \\
 & $\alpha_{\parallel}$ & -0.001 & 0.007 & 0.093 & 0.032 \\
 
 \hline
 
\end{tabular}
\caption{Statistics from runs at different values of $k_{\text{fg}}$ in Fig.~\ref{kfg-dist}. Catastrophic fits removed.}
\label{tab-kfg}
\end{table}

\subsubsection{Foreground cuts}
\label{sec:fgcuts}

In Fig.~\ref{kfg-dist} and Table~\ref{tab-kfg} we show the results of fitting to 5000 simulations generated under fiducial conditions but with varying values of the foreground cut, $k_{\rm fg}$.

The variance of the recovered $\alpha_{\parallel}$ and $\alpha_{\perp}$ distributions is similar for each value of $k_{\text{fg}}$. For $\alpha_{\parallel}$, the distribution is unbiased and appears approximately Gaussian within 10\% of the fiducial value, but has an enhanced tail into the $\alpha_{\parallel} > 1$ region. The $\alpha_{\perp}$ parameter has a bias that changes over the range of $k_{\text{fg}}$ values however, being biased low at small $k_{\text{fg}}$ and then high at $k_{\text{fg}} = 0.05$ Mpc$^{-1}$. 
We plot the median bias as a function of $k_{\text{fg}}$ in the lower panel of Fig.~\ref{kfg-dist}.

While $\alpha_{\parallel}$ is recovered with a somewhat large variance in these runs due to the measurement being sample variance-limited, it is at least robust to the foreground cut value. The $\alpha_{\perp}$ parameter becomes even more difficult to recover at higher foreground cut values however. In the correlation function model, there is change to the shape of the BAO feature under different foreground cuts that was visible in Fig.~\ref{models}. This seems to negatively impact the prospects for recovering $\alpha_\perp$, and may point to a need to use a more sophisticated forward model.

\subsubsection{Non-optimal covariance}
\label{sec:nonoptcov}

\begin{figure}
	\includegraphics[scale=0.23]{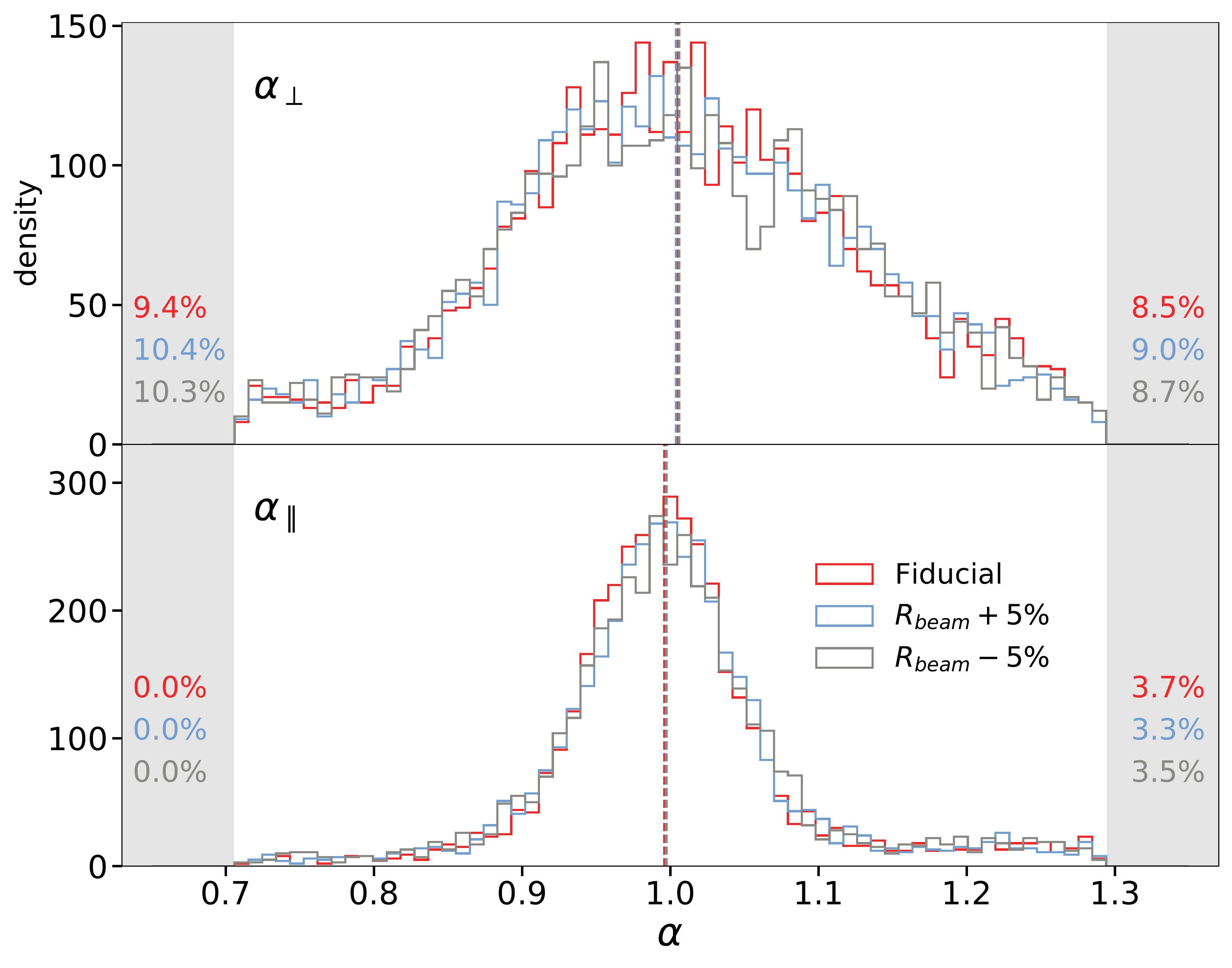}
	\caption{Recovered $\alpha$ parameters from fits to 5000 realisations generated using a covariance matrix with $R_{\text{beam,cov}} +/- 5\%$. No significant differences in recovery are seen as a result of the shift in covariance beam parameter. }
	\label{cov-dist}
\end{figure}

Fig.~\ref{cov-dist} and Table~\ref{tab-cov} show the results of 5000 fits under fiducial conditions, but now changing the value of the Gaussian beam width in the covariance matrix only. The value of $R_{\rm beam}$ used to compute the mean model (the correlation function) is left unchanged.

The distributions at $R_{\text{beam,cov}} +5\%$ and $R_{\text{beam,cov}} -5\%$ show only small differences with the fiducial case. The median $\alpha_{\perp}$ values for these two case are 0.1$\%$ and $0.2\%$ larger than the fiducial run, but aside from this the runs share almost identical statistics. This suggests that small model errors in the calculation of the covariance matrix should not significantly bias the recovery of the $\alpha$ parameters from the correlation function.

\subsection{MCMC analysis of MeerKAT-like data}

In this section, we perform an illustrative MCMC analysis of a single Gaussian random realisation from the fiducial case, including all of the anisotropic effects, and the standard assumptions for the noise level and survey area at $z = 0.3915$.

Fig.~\ref{fig:mcmc} shows the posterior distribution for all of the free fitting model parameters after using {\tt emcee} \citep{2013PASP..125..306F} with the Gaussian likelihood for the correlation function multipoles from Eq.~\ref{eq:likelihood}, and uniform priors on the parameters. The true (input) values of relevant parameters are shown as blue lines and points.

From Fig.~\ref{fig:mcmc}, we can see that the marginal posterior distributions for $\alpha_\perp$ and $\alpha_\parallel$ have widths that are essentially consistent with the standard deviation computed for the distribution over 5,000 random realisations of the data (see Table~\ref{tab-contribs}, final two lines). For this realisation, the best-fit $\alpha$ parameters are $\alpha_{\perp} = 0.942^{+0.090}_{-0.101}$ and $\alpha_{\parallel} = 1.052^{+0.046}_{-0.056}$ (68\% CL), to be compared with ensemble standard deviations of $\sigma(\alpha_\perp) = 0.12$ and $\sigma(\alpha_\parallel) = 0.09$ from Table~\ref{tab-contribs}, which includes the influence of the non-Gaussian tails. Importantly, there are no strong correlations between the $\alpha_{\perp}$ and $\alpha_{\parallel}$ parameters and the continuum fitting polynomial coefficients. 
While not evident in this particular case, we have observed that the Gaussian beam width parameter $R_{\text{beam}}$ can interact strongly with the continuum parameters, allowing a substantial probability mass to appear away from the true $\alpha$ values. This motivated us to choose the relatively narrow prior range on $R_{\text{beam}}$ for the least squares fitting runs in the previous sections. 

\section{Conclusions}
\label{sec:conclusions}

The baryon acoustic oscillation (BAO) scale, acting as a statistical standard ruler, contains valuable information about the angular diameter distance and cosmic expansion rate as a function of redshift. Detecting and measuring the BAO feature in the redshift-space correlation function will be an important validation step for the 21cm IM method, as it presents an unambiguous and well-understood target that is difficult to mask or mimic with systematic effects. This is in contrast to (e.g.) the broadband shape of the power spectrum, which can be strongly affected by errors in modelling the effects of the instrumental beam and the removal of bright foreground contamination for example.

While recovery of the BAO feature from the 21cm signal may proceed in either the Fourier or real domain \citep[e.g.][]{Chang2007, Bull2015, Seo2016, Villaescusa-Navarro2017, Soares2021}, we have chosen to focus on the real domain here as we believe it has some advantages for a conservative first analysis leading to a detection with an autocorrelation-type IM experiment. In particular, a correlation function measurement can be performed directly on the intensity maps, without needing to first Fourier transform the data. Fourier transforming risks introducing ringing and mode-coupling artifacts (e.g. due to the RFI mask) that can swamp the signal given the large dynamic range between the cosmological 21cm signal and the foregrounds. While these effects can be mitigated in a Fourier analysis \citep[e.g.][]{2019MNRAS.484.2866O, 2021MNRAS.500.5195E}, is is useful to be able to sidestep them as a way of simplifying analyses. The cost of this approach is that the correlation function and its covariance are harder to model and compute. 

\begin{table}
\centering
\begin{tabular}{ |c|c|c|c|c|c|  }
 \hline
 Covariance & & $\Delta$Med. &  $\Delta$Mean & $\sigma$ & MAD\\
 \hline

 Fiducial & $ \alpha_{\perp}$ & 0.005 & 0.010 & 0.118 & 0.078 \\
 & $ \alpha_{\parallel}$ & -0.004 & 0.002 & 0.081 & 0.036 \\\hline
 $R_{\text{beam,cov}} + 5\%$& $ \alpha_{\perp}$ & 0.004 & 0.008 & 0.115 & 0.082 \\ 
 & $ \alpha_{\parallel}$ & -0.003 & 0.002 & 0.080 & 0.038 \\ \hline
 $R_{\text{beam,cov}} - 5\%$& $ \alpha_{\perp}$ & 0.006 & 0.008 & 0.121 & 0.086 \\ 
 & $ \alpha_{\parallel}$ & -0.003 & 0.003 & 0.086 & 0.040 \\
 
 \hline
\end{tabular}
\caption{Statistics from runs varying $R_{\text{beam,cov}}$ in Fig.~\ref{cov-dist}. Catastrophic fits removed.}
\label{tab-cov}
\end{table}

\begin{figure*}
	\includegraphics[width=2\columnwidth]{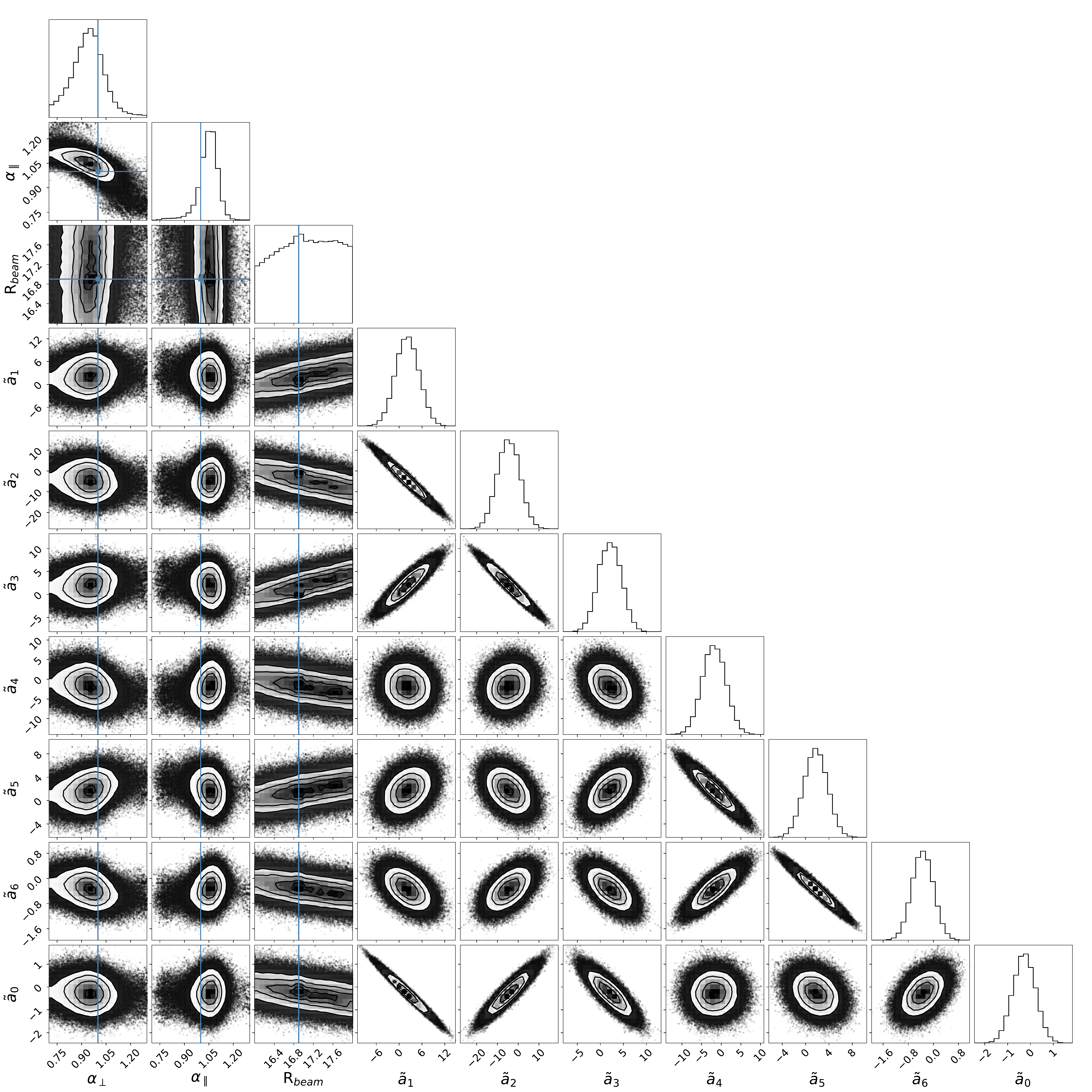}
	\caption{MCMC fit to a single random realisation of the correlation function multipoles under fiducial MeerKAT conditions (including RSDs, a Gaussian beam model, and a $k_{\rm fg} = 0.01$ Mpc$^{-1}$ foreground cut) at $z = 0.3915$. Continuum parameters have been rescaled for clarity, so that $\tilde a_n = a_n (r/r_m)^{p} \times 1000$, where $r_m = 50~{\rm Mpc}$ is an arbitrary reference scale and $p$ is the appropriate exponent from Eq.~\ref{eq:continuum}. True values of the input parameters $\alpha_{\perp},\alpha_{\perp},R_{\text{beam}}$ are shown with vertical lines. We recover the input value of $R_{\text{beam}}$ with the fit to this realisation, but recovered $\alpha$ values are biased at the 5\% level, in line with results obtained from our least-squares fitting runs.}
	\label{fig:mcmc}
\end{figure*}

In this paper, we have constructed an analytic model of the redshift-space 21cm correlation function, its multipoles, and their covariance, all in the presence of several key anisotropic systematic effects. These are: the angular smoothing effect due to the instrumental beam; redshift-space distortions; and the removal of radial Fourier modes due to foreground filtering. Each of these effects changes the correlation structure of the covariance matrix, and either suppresses or masks the radial or transverse BAO feature to some extent. We have then demonstrated how the radial and transverse BAO scales (denoted by the radial and transverse shift parameters, $\alpha_\parallel$ and $\alpha_\perp$) can be successfully extracted in the presence of these complications for a realistic 21cm autocorrelation survey with a similar configuration to the MeerKLASS L-band survey on MeerKAT (covering $0 \lesssim z \lesssim 0.46$). Our analysis is based on applying least-squares fits of a phenomenological correlation function model to ensembles of thousands of Gaussian random realisations of the binned multipoles of the 21cm correlation function, with noise properties calculated according to the relevant analytic covariance matrix model. 

As found by previous authors \citep[e.g.][]{Villaescusa-Navarro2017}, the relatively low angular resolution of the MeerKAT dishes at the relevant frequencies results in a BAO feature that is considerably smoothed in the transverse direction, while remaining well-defined along the line of sight. Simply performing a spherical average of the correlation function results in a washed-out, and possibly undetectable, BAO feature, and so an anisotropic analysis is required to maximise the amount of information that can be recovered. We use a Legendre multipole expansion of the correlation function for this. Other effects, such as RSDs and the $k_{\parallel}$ foreground cut, can also enhance the smoothing effect and affect the shape and normalisation of the correlation function multipoles, but the instrumental beam angular resolution effect is the dominant cause of the smearing of the BAO scale. When a multipole analysis is implemented, we find that the BAO feature remains well-defined in the quadrupole even when it has been smoothed away completely in the monopole, despite the quadrupole measurements being noisier.

The anisotropic effects also affect the correlation structure of the multipole covariance matrix. By far the largest effect is due to the RSDs, which introduce an anti-correlation between the monopole and quadrupole. When including the beam response and a foreground cut, the width of the band along the diagonal of the matrix is increased, denoting enhanced correlations of $\xi_\ell(r)$ between neighbouring separation bins, especially on smaller scales (lower values of $r$). The variance (along the diagonal of the covariance matrix) is reduced due to the smoothing effect of the beam on the 21cm fluctuation field, but this does not result in reduced uncertainties (e.g. when evaluating the likelihood function) when the enhanced correlations between bins are taken into account.

A moderate anti-correlation is also introduced into the off-diagonal region of the monopole block of the covariance matrix when the beam and foreground cut are added (see Fig.~\ref{corrcovs}). This is largely due to the non-local effect of the foreground cut, which largely affects the overall normalisation of the correlation function multipoles (e.g. see Fig.~\ref{models}), thus coupling bins at high and low values of $r$.
In all cases, the increase in correlated uncertainties has a detrimental effect on the ability to recover the BAO $\alpha$ parameters from the simulations; by correlating neighbouring separation bins, we lose our ability to sharply resolve the BAO feature, i.e. there is an effective loss in resolution as a function of separation $r$.

To model the retrieval of the BAO scale in a semi-realistic setting, we performed least-squares fits of a phenomenological correlation function model to several thousand Gaussian random realisations of the correlation function multipoles based on our analytic covariance matrix calculations. The recovered distribution for the $\alpha_{\perp}$ parameter is much wider than the $\alpha_{\parallel}$ distribution when the beam smoothing effect is included, reflecting the loss of angular information. By performing a multipole analysis, we are able to avoid the total loss of the BAO feature due to beam smoothing that led \cite{Villaescusa-Navarro2017} to propose the line-of-sight power spectrum, $P_{1{\rm D}}(k_\parallel)$, as an alternative statistic to the (spherically-averaged) correlation function. A particular advantage of the correlation function multipole analysis is that it retains angular information when it is available, for example at lower redshifts where the transverse BAO feature is not completely smoothed out. This is in contrast to the $P_{1{\rm D}}(k_\parallel)$ analysis, which proactively averages away all transverse information.

The distribution of recovered $\alpha_{\perp}$ values is typically slightly non-Gaussian, with a larger tail into the $\alpha_{\perp} > 1$ region. The skewness of the distribution is enhanced when approximate Gaussian beam models are used during the fitting process instead of the more accurate MeerKAT beam, as shown in Figs.~\ref{MKvG-dist} and \ref{MKvG-dist2}, resulting in a biased recovery of $\alpha_{\perp}$ on average. This bias can be reduced by carefully matching the FWHM of the Hankel transform of the Gaussian beam to that of the MeerKAT beam, rather than performing the matching in real space. The bias on the $\alpha_\perp$ parameter is then decreased by approximately 30\%. Conversely, $\alpha_{\parallel}$ is not significantly biased by the choice of an incorrect/approximate beam model in any case.

In the fiducial (MeerKLASS survey) scenario, the distribution of recovered $\alpha_\parallel$ values is slightly broadened by the addition of the beam smoothing and foreground cuts (Fig.~\ref{contribs-dist}), but to a far lesser extent than for $\alpha_\perp$. Small but non-negligible tails are observed in the distribution at both high and low values of $\alpha_\parallel$ even for the base cosmology-only case. This suggests that some realisations of the correlation functions, by chance, exhibit features that are harder to disentangle from (e.g.) the continuum fitting parameters, leading to spurious correlations that bias the recovery of $\alpha_\parallel$ in some cases. This is to be expected when there are substantial correlations between neighbouring $r$ bins, which will tend to produce occasional random realisations that are more smoothed-out (and thus continuum-like) than the underlying mean correlation function. A mild manifestation of this effect is visible in the posterior distribution of the fitting parameters from the MCMC analysis that we performed on a single random realisation of the correlation function multipoles (see Fig.~\ref{fig:mcmc}). In this case, it can be seen that the $R_{\text{beam}}$ parameter is correlated with several of the continuum fitting parameters. Stronger manifestations of this effect are the likely cause of the heavier tails in the $\alpha_\parallel$ distribution.

There is a small effect on the recovered BAO scale distributions as increasingly severe foreground cuts are applied (Fig.~\ref{kfg-dist}). For $\alpha_\perp$, changing the foreground cut results in a changing bias -- up to $\pm 2\%$ in the most extreme cases -- with a sign that changes from negative to positive as $k_{\rm fg}$ increases. The bias on $\alpha_\parallel$ is essentially negligible however, with the main effect of changing $k_{\rm fg}$ being to slightly modify the variance of the distribution. In fact, the only effect that results in large changes in the $\alpha_\parallel$ distribution is when the survey area is increased to an SKAO-like value of $f_{\rm sky} = 0.5$ (Fig.~\ref{noise-dist}). This greatly reduces the sample variance, producing a narrower $\alpha_\parallel$ distribution (going from $\sigma \approx 8\%$ to $3\%$) and strongly suppressing the non-Gaussian tails and catastrophic outliers that are observed for smaller values of $f_{\rm sky}$. Changing $f_{\rm sky}$ also has a large effect on the $\alpha_\perp$ distribution, reducing its width from $\sigma \approx 12\%$ to $7\%$, but still leaving substantial non-Gaussianity.

Finally, we note that our results are not particularly sensitive to approximations made in the analytic covariance matrix calculation. Recovery of both $\alpha$ parameters was unaffected by a 5\% level error in the value of $R_{\rm beam}$ when calculating the covariance matrix for example (Fig.~\ref{cov-dist}).

Taken together, our results demonstrate that the radial BAO parameter, $\alpha_\parallel$, can be recovered robustly from a correlation function multipole analysis with 21cm autocorrelation data, even in the presence of severe anisotropic systematic effects. The same is not true of the transverse BAO parameter, $\alpha_\perp$, although if sufficient care is taken with (e.g.) the modelling of the beams, useful information can still be recovered, and there is no need to completely average away transverse Fourier modes, as suggested by \cite{Villaescusa-Navarro2017}.



    
To conclude, we highlight some of the limitations of our analysis. An analytic analysis of this nature is inherently simplified, but provides us with a means to build up a picture of how the BAO recovery process is likely to operate in many different scenarios. In particular, our reduction of the foreground cleaning process to a threshold excision of smaller line-of-sight modes is quite simplistic. In our analysis, we have found that a line-of-sight mode only analysis should be unbiased, but this picture may change if the more complex interactions between foreground removal algorithms and the beam response function are considered, as in \cite{Matshawule2020}. Our method has also made exclusive use of the linear matter power spectrum, therefore ignoring non-linear corrections. As such, the effects of various treatments of non-linearities, including the potential for performing BAO reconstruction \citep{Obuljen2016, Seo2016}, have not been considered.

More direct simulations that produce and analyse 3D datacubes of the 21cm brightness temperature field itself, rather than only the correlation function, would allow for more realistic treatments of these effects, despite being more computationally intensive (e.g. see \cite{Cunnington2019}; Vos Gin\'es et al., {\it in prep.}), complementing the partially-analytic correlation function and covariance calculations we have used here.

{\it Note added:} During the late stages of preparation of this paper, we were made aware of an independent project to calculate beam convolution effects on the 21cm correlation function (Vos Gin\'es et al., {\it in prep.}). This uses a suite of numerical simulations, instead of an analytic calculation like the one we have presented here. A preliminary comparison suggests good qualitative agreement between the two approaches.

\vspace{-2em}
\section*{Acknowledgements}

We are grateful to S.~Avila, C.~Blake, P.~Carrilho, S.~Choudhuri, S.~Cunnington, J.~Fonseca, H.~Garsden,  G.~Jelic-Cizmek, A.~Pourtsidou, M.~Santos, P.~Soares, and B.~Vos Gin\'es for useful comments and discussions.
FK acknowledges support from an STFC PhD studentship. PB acknowledges funding for part of this research from the European Research Council (ERC) under the European Union's Horizon 2020 research and innovation programme (Grant agreement No. 948764), and from STFC Grant ST/T000341/1. We acknowledge use of the following software: {\tt emcee} \citep{2013PASP..125..306F}, {\tt matplotlib} \citep{matplotlib}, {\tt numpy} \citep{numpy}, {\tt pyfftlog} \citep{pyfftlog}, and {\tt scipy} \citep{2020SciPy-NMeth}.

\section*{Data Availability}

The Python code used to produce the results in this paper is available from \url{https://github.com/fraserlkennedy/21cmCorrelationFn}.



\bibliographystyle{mnras}
\bibliography{meerkat_bao}


\appendix

\section{Fast integrals with FFTLog}
\label{app:fftlog}
Carrying out the integrals $I_\ell(r)$ from Sect.~\ref{sec:corrfn} is numerically challenging, as the spherical Bessel function $j_\ell(x)$ oscillates rapidly for large values of the argument. Rather than direct integration, we make use of {\tt FFTLog} \citep{Talman1978, Hamilton1999} to carry out the integral. This method is applicable to general Hankel transforms, so we need only exchange the spherical Bessel function $j_\ell(x)$ for a Bessel function of the first kind $J_\ell(x)$ in our model to carry out the process. {\tt FFTLog} works by noting that, when switching to a logarithmic scale in the independent variable and assuming a logarithmic period $L$, i.e.
\begin{equation}
f(r^\prime) = f(r e^L),
\end{equation}
that the general Hankel transform 
\begin{equation}
    A(r) = \int k J_{\ell}(kr) \tilde A(k) dk
\end{equation}
then takes the form of a convolution
\begin{equation}
    A(\ln r) = \int e^{\ln k + \ln r} J_{\ell}(\ln k + \ln r) \tilde A(\ln k) d(\ln k).
\end{equation}
In these circumstances, it is possible to evaluate the entire convolution integral by Fourier transforming the individual terms, multiplying them together, and then performing the inverse Fourier transform. Computation time is greatly decreased by avoiding direct integration in this way. In our calculations, we use the {\tt pyfftlog} package \citep{pyfftlog}, which has additional functionality aimed at mitigating the susceptibility of both FFT steps to ringing.

\section{Derivation of the multipole covariance matrix}
\label{sec:covderiv}
In this appendix, we derive an analytic expression for the multipole covariance matrix under the assumption of Gaussianity of the 21cm correlation function. Our derivation follows the method and conventions of \cite{Tansella2018}.

The observed correlation function $X(\mathbf{r})$ for voxels separated by comoving vector $\mathbf{r}$ can be written as
\begin{equation}
X(\mathbf{r}) \equiv \langle(\delta_i(\mathbf{x}) + n_{i})(\delta_{j}(\mathbf{x}+\mathbf{r}) + n_{j})\rangle,
\end{equation}
where $i,j$ label the voxels, $n_i$ is a shot noise term, and the angle brackets denote spatial averaging, which is equivalent to an ensemble average if the ergodic theorem applies \citep{Peebles1980}. The covariance of the measured correlation function values in bins of separation $\mathbf{r}$ and $\mathbf{r^\prime}$ is then
\begin{equation}
\mathbf{C}(\mathbf{r},\mathbf{r^\prime}) = \langle X(\mathbf{r}) X(\mathbf{r^\prime}) \rangle - \langle X(\mathbf{r}) \rangle\langle X(\mathbf{r^\prime}) \rangle.
\end{equation}
Next, we expand the expression above, labeling voxel positions with indices $(i, j, k, l) = (\vec{x},\, \vec{x}+\vec{r},\, \vec{x}^\prime,\, \vec{x}^\prime+\vec{r}^\prime)$. A set of trispectra and products of two-point functions results. Those with odd numbers of $\delta$ and $n$ terms (e.g. $\langle\delta\delta\delta n\rangle$) drop out, since the noise $n$ is assumed to be uncorrelated with the density field. We denote the two-point terms for the signal and noise as $\langle\delta_i\delta_j\rangle = \xi_{ij}$ and $ \langle n_in_j\rangle = N_{ij}$ respectively, and re-express the outer expectation value operation as an integral over the spatial domains of $\mathbf{r}$ and $\mathbf{r^\prime}$ to obtain
\begin{dmath}
\mathbf{C}(\mathbf{r}, \mathbf{r^\prime}) = \frac{1}{V^2}\int_{V\times V} d^3\mathbf{x} d^3\mathbf{x^\prime} \bigg[\xi_{ik} \xi_{jl} + \xi_{il} \xi_{jk} + \xi_{ik} N_{jl} + \xi_{il} N_{jk} + N_{ik} \xi_{jl} + N_{il} \xi_{jk} + N_{ik} N_{jl} + N_{il} N_{jk}\bigg].
\end{dmath}
We next assume the noise covariance to be diagonal (uncorrelated), $N_{ij} = \bar{n}^{-2} \delta_{ij}$, where $\delta_{ij}$ is the Kronecker delta function. Inserting this into the expression above and re-expressing the terms as functions of position/separation, we obtain \cite{Tansella2018}: 
\begin{dmath}
	C(\mathbf{r},\mathbf{r^\prime}) = \frac{1}{V^2}\int_{V\times V} d^3\mathbf{x}\, d^3\mathbf{x^\prime} \big[\xi(\mathbf{x}-\mathbf{x^\prime})\,\xi(\mathbf{x}+\mathbf{r}-\mathbf{x^\prime}-\mathbf{r^\prime}) \\
	~~~~~~~+  \xi(\mathbf{x}+\mathbf{r}-\mathbf{x^\prime})\,\xi(\mathbf{x}-\mathbf{x^\prime}-\mathbf{r^\prime})\big] \\
	+ \frac{2}{V\bar n} \big[\xi(\mathbf{r}-\mathbf{r^\prime})+\xi(\mathbf{r}+\mathbf{r^\prime})\big] \\
	+ \frac{1}{\bar n^2 }\big[ \delta^{(3)}(\mathbf{r}-\mathbf{r^\prime})+\delta^{(3)}(\mathbf{r}+\mathbf{r^\prime}) \big].
\end{dmath}
The terms involving products of $\xi$ are convolutions, made plainer after substitution for $\mathbf{x}-\mathbf{x^\prime}$. These can be expressed more simply in harmonic space, where we obtain
\begin{dmath}
	C(\mathbf{r,r^\prime}) = \frac{1}{V(2\pi^3)}\int_{V}d^3k\bigg[P^2(\mathbf{k}) + \frac{2}{\bar n}P(\mathbf{k}) + \frac{1}{\bar n^2}\bigg]\\
	~~~~~~~~~~~~~~~~~\times \bigg(  e^{i\mathbf{k}\cdot (\mathbf{r-r^\prime})} + e^{i\mathbf{k}\cdot (\mathbf{r+r^\prime})} \bigg).
\end{dmath}
Next, we apply the plane wave expansion for the exponentials,
\begin{dmath}
	\exp(i\mathbf{k\cdot r}) = \sum_{\ell = 0}^{\infty} i^\ell (2\ell+1) \mathcal{P}_\ell(\mu) j_\ell(kr),  	
\end{dmath}
where $\mathcal{P}_\ell $ and $j_\ell$ are the Legendre and spherical Bessel functions of degree $\ell$, and $\mu = \mathbf{\hat k \cdot \hat r}$. 
We then obtain
\begin{dmath}
	C(\mathbf{r,r^\prime}) = \frac{1}{V(2\pi^3)}\int_{V}d^3k\bigg[P^2(\mathbf{k}) + \frac{2}{\bar n}P(\mathbf{k}) + \frac{1}{\bar n^2}\bigg] \\
	~~~~~~~\times 
	\sum_{\ell,\ell^\prime} \big[i^{\ell+\ell^\prime} + i^{\ell+\ell^\prime}\big](2\ell+1)(2\ell^\prime+1) \\
	~~~~~~~~~~~\times \mathcal{P}_\ell(\mu) 	\mathcal{P}_{\ell^\prime}(\mu^\prime) j_\ell(kr)j_{\ell^\prime}(-kr^\prime),
\end{dmath}
where the sum $i^{\ell+\ell^\prime} + i^{\ell+\ell^\prime}$ has been left for clarity; using $j_{\ell^\prime}(-kr^\prime) = -1^{\ell^\prime} j_{\ell^\prime}(kr^\prime)$, the sum becomes $i^{\ell+\ell^\prime} + i^{\ell-\ell^\prime}$. Note that $P(\mathbf{k})$ is an even function, which implies that the power spectrum terms are only non-vanishing when $\ell,\ell^\prime$ are even. Thus, a sum and difference of even powers of $i$ will always return the same answer, and the two terms can be collected. We then obtain
\begin{dmath}
	C(\mathbf{r,r^\prime}) = \frac{1}{V(2\pi^3)}\int_{V}d^3k\bigg[P^2(\mathbf{k}) + \frac{2}{\bar n}P(\mathbf{k}) + \frac{1}{\bar n^2}\bigg] \\
	~~~~~~~~~~~~~~~ \times 
	\sum_{\ell,\ell^\prime} 2i^{\ell-\ell^\prime}(2\ell+1)(2\ell^\prime+1) \\
	~~~~~~~~~~~~~~~ \times \mathcal{P}_\ell(\mu) 	\mathcal{P}_{\ell^\prime}(\mu^\prime) j_\ell(kr)j_{\ell^\prime}(kr^\prime).
\end{dmath}
With this expression in hand, we can now expand the various anisotropic factors that multiply the power spectrum, e.g. due to the beams and foreground removal.
Assuming axisymmetry, the multipole expansion of the product of the isotropic cosmological spectrum and the anisotropic modulation can be expressed
\begin{dmath}
	P(\mathbf{k}) = F(\mathbf{k})P(k) = P(k) \sum_{\ell}  c_\ell^{(1)}(k) \mathcal{P}_\ell(\nu),
\end{dmath}
using the $c_\ell^{(n)}$ notation defined in Section~\ref{sec:corrfn}. The expansion for the squared term is also needed:
\begin{dmath}
 P^2(\mathbf{k}) = 	F^2(\mathbf{k})P^2(k) = P^2(k)\sum_{\ell} c_\ell^{(2)}(k)  \mathcal{P}_\ell(\nu) 
\end{dmath}
Indices for both expansions may be run along the same index $L$ as such:
\begin{dmath}
	C(\mathbf{r,r^\prime}) = \frac{1}{V(2\pi^3)}\int_{V}d^3k \\
	\bigg[\sum_L \bigg( P^2(k)c_L^{(2)}(k) + \frac{2}{\bar n}P(k)c_L^{(1)}(k)\bigg)\mathcal{P}_{L}(\nu)  + \frac{1}{\bar n^2}\bigg] \\
	\times \sum_{\ell,\ell^\prime} 2i^{\ell-\ell^\prime}(2\ell+1)(2\ell^\prime+1) \mathcal{P}_\ell(\mu) 	\mathcal{P}_{\ell^\prime}(\mu^\prime) j_\ell(kr)j_{\ell^\prime}(kr^\prime) 	
\end{dmath}

\balance

The angular part of the integral now has the form
\begin{dmath}
	\int d\Omega_{\mathbf{k}} \mathcal{P}_\ell (\mu) \mathcal{P}_{\ell^\prime} (\mu)\mathcal{P}_L(\nu) \\
	= \mathcal{W}^{L\ell\ell^\prime}_{000} \sqrt{ \frac{(4\pi)^5}{(2L+1)(2\ell+1)(2\ell^\prime+1)}  } \\
	\times \sum_{M,m, m^\prime} \mathcal{W}^{L\ell\ell^\prime}_{Mm m^\prime} Y_{LM}^*(\mathbf{\hat n})\, Y_{\ell m}^*(\mathbf{\hat r})\, Y_{\ell^\prime m^\prime}^*(\mathbf{\hat{r}^\prime})
\end{dmath}
where we have applied the addition theorem to each Legendre polynomial 
\begin{equation}
\mathcal{P}_\ell(\mu) = \mathcal{P}_\ell(\mathbf{\hat k \cdot \hat r} ) = \frac{4\pi}{2\ell+1} \sum_{m=-\ell}^\ell Y_{\ell m}(\mathbf{\hat k}) Y_{\ell m}^*(\mathbf{\hat r}) 
\end{equation}
and used this integral over 3x product spherical harmonic identity, $\mathcal{W}$ denoting the Wigner-3j symbol,
\begin{dmath}
\int d\Omega_{\mathbf{k}} Y_{LM}(\mathbf{\hat k}) Y_{\ell m}(\mathbf{\hat k}) Y_{\ell^\prime m^\prime}(\mathbf{\hat k})  \\
= \sqrt{ \frac{(2L+1)(2\ell+1)(2\ell^\prime+1)}{4\pi}  }\mathcal{W}^{L\ell\ell^\prime}_{000}\mathcal{W}^{L\ell\ell^\prime}_{Mmm^\prime}.
\end{dmath}
The LOS normal is chosen as $\mathbf{\hat{n}} = \mathbf{e_z}$ which sets the index $M = 0$ via $Y_{LM}(\mathbf{ \hat n}) = \sqrt{\frac{2L+1}{4\pi}}\delta_{M,0}$ and the 3D covariance, before multipoles are taken is as seen in Eq.~A.22 of \cite{Tansella2018}, but now includes anisotropic factors which are functions of $k$ in general
\begin{dmath}
	C(\mathbf{r,r^\prime}) = \frac{1}{V\pi} \sum_{L,\ell, \ell^\prime, m, m^\prime} i^{\ell-\ell^\prime} \sqrt{(2\ell+1)(2\ell^\prime+1)} \\ \mathcal{W}^{L\ell\ell^\prime}_{000}\mathcal{W}^{L\ell\ell^\prime}_{0mm^\prime} Y_{\ell m}^*(\mathbf{\hat r}) Y_{\ell^\prime m^\prime}^*(\mathbf{\hat r^\prime}) \\ \times 
	\int dk k^2 \bigg[P^2(k)c_L^{(2)}(k) + \frac{2}{\bar n} P(k)c_L^{(1)}(k) +\delta_{0,L} \frac{1}{\bar n^2}   \bigg]  j_\ell(kr)j_{\ell^\prime}(kr^\prime)
\end{dmath}
In order to calculate the multipoles of this expression in both angular coordinates, the spherical harmonics are converted back to Legendre polynomials via $\mathcal{P}_n(\nu) = \sqrt{\frac{4\pi}{2n+1}} Y_{n0}(\mathbf{\hat r})$. This sets all $m, m^\prime = 0$, and the multipoles of the 3D covariance can be evaluated
\begin{equation}
    \text{C}_{\ell\ell^\prime}(r_i,r_j) = \frac{(2\ell+1)(2\ell^\prime+1)}{4} \int_{-1}^1 d\mu \int_{-1}^1 d\mu^\prime \mathcal{P}_\ell(\mu)\mathcal{P}_{\ell^\prime}(\mu^\prime)C(\mathbf{r,r^\prime})
\end{equation}
After which only summation over $L$ remains, which we exchange for $n$ in the final expression. The pure shot noise term can be further simplified by using the orthogonality of the spherical Bessel function, 
\begin{dmath}
	\int_0^{\infty} dk k^2 j_\ell(kr)j_{\ell^\prime}(kr^\prime) = \delta^{(1)}(r-r^\prime) \frac{\pi}{2r^2}.
\end{dmath}
We are left with the final result:
\begin{dmath}
    \text{C}_{\ell\ell^\prime}(r_i,r_j) =  \frac{i^{\ell-\ell^\prime}}{V\pi^2}
    \\ \times \Bigg(\frac{(2\ell+1)\pi}{2\bar n^2 L_p r^2} \delta_{ij} \delta_{\ell\ell^\prime} + \frac{2}{\bar n}A_{\ell\ell^\prime}(r_i,r_j) + B_{\ell\ell^\prime}(r_i,r_j) \Bigg)
\end{dmath}
which has used the prescription that $\delta(r-r^\prime) = \frac{\delta_{r,r^\prime}}{L_p}$, with $L_p$ being the covariance pixel size and the $\delta_{\ell\ell'}$ appearing after evaluation of the Wigner-3j symbol on the diagonal. Functions $A$ and $B$ are defined immediately following Eqn.~\ref{eq:corrfn_multipoles}, which we reproduce here for convenience:
\begin{align}
    A_{\ell\ell^\prime}(r_i,r_j)& =  ~(2\ell+1)(2\ell^\prime+1)\nonumber \\
    & \times \int_0^{\infty}dk k^2 P(k) j_{\ell}(kr_i) j_{\ell^\prime}(r_j) \sum_{n} c^{(1)}_n(k) \bigg(\mathcal{W}^{\ell\ell^\prime n}_{000}\bigg)^2 \nonumber
\end{align}
\begin{align}
    B_{\ell\ell^\prime}(r_i,r_j)& = ~(2\ell+1)(2\ell^\prime+1)  \nonumber \\
    & \times \int_0^{\infty}dk k^2 P^2(k) j_{\ell}(kr_i) j_{\ell^\prime}(r_j) \sum_{n}c^{(2)}_n(k) \bigg(\mathcal{W}^{\ell\ell^\prime n}_{000}\bigg)^2 \nonumber.
\end{align}
Critically, the terms $c_n$ are now under the integral signs, in contrast with the result in \citet{Tansella2018}, where only cases where they were multiplicative constants were considered. 

\bsp	
\label{lastpage}
\end{document}